\documentclass[12pt,one column,showpacs,preprintnumbers,pra,aps]{revtex4}
\usepackage{amsmath}
\usepackage{amsfonts}
\usepackage{graphicx}
\usepackage{dcolumn}
\usepackage{array}
\usepackage{bm}

\begin{document}
\title{An extended antiferromagnetic model for quasielectrons in identical ionic-covalent bonds}
\author{ C. F. Huang}
\affiliation{ 2nd Patent Division, Intellectual Property Office, Ministry of Economic Affairs, Taipei, Taiwan 106, R.O.C.}

\begin{abstract}
The antiferromagnetic (AF) model is generalized for the quasielectron system composed of identical ionic-covalent dimers. The density-fluctuation and covalent-correlation operators are constructed based on the extended AF density matrices, and the quasielectron system is decomposed into 4-level subsystems for the electron ionization and affinity. By considering the nearest-neighbor hopping near the covalent limit, we can see the importance of the bonding coefficients to the effective mass of the excited carrier in the crystal of the zincblende structure. 
\end{abstract}
\pacs{ }
\maketitle
\section{Introduction}
Different quasiparticles are taken into account for many-electron systems. Based on the Bardeen Cooper Schrieffer (BCS) theory \cite{James}, the Bogoliubov-BCS quasiparticles are constructed under the superconducting (SC) order by considering Bogoliubov-deGennes or Hartree-Bogoliubov equations \cite{Ghosal,Barankov,Paar}. The antiferromagnetic (AF)  quasielectrons are introduced when the orbital antiferromagnetism or the d-density wave (DDW) order becomes important \cite{Gerami,Chakravarty}, and the quasiparticles under multiple orders are discussed in the literature \cite{Bena,Kee,Huang1,Laughlin,Ramshaw}. The Hubbard model \cite{Mahan} may help us to clarify the SC and AF behaviors, and the 4-level Hubbard dimer \cite{Matlak1,Matlak2} is constructed by including the up- and down-spin orbitals at the two sites. The ionic-covalent chemical bond can be approximated by such a dimer when the two sites correspond to the atomic orbitals, and we have the Heitler-London state \cite{Fulde,Soos}, which is denoted by $| \Psi _{c} \rangle$ in this manuscript, for the covalent limit. The ionic state $| \Psi _{i} \rangle$ may become dominant in the hetero-diatom bond, in which we shall consider different on-site energies introduced in the ionic Hubbard model \cite{Kampf,Buzatu}. 

The Bloch states \cite{Fulde,Grosso,Grafenstein} are important to construct the quasiparticle orbitals, including the plane-wave ones for the nearly free carriers in crystals. Such states can be introduced based on the mean-field methods such as the Hartree-Fock (HF) approach, which yields the self-consistent-field (SCF) solutions \cite{Fulde,Grafenstein,Stoyanova}. It has been discussed in the literature how to include the correlation energy beyond the HF approach by considering the coupled-cluster corrections such as those due to coupled-cluster doubles (CCDs) \cite{Fulde,Grafenstein,Stoyanova,Doll,Talukdar}. In addition, my group \cite{Huang1} has used the AF part of the extended Bogoliubov-BCS quasiparticles to construct the energy form of the one-bond system by considering \cite{Prasad}
\begin{eqnarray} 
| \Psi _{b} \rangle = \alpha _{i} | \Psi _{i} \rangle + \alpha _{c} | \Psi _{c} \rangle.
\end{eqnarray}
Here $| \Psi _{b} \rangle$ denotes the bonding wavefunction of the half-filled ionic-covalent bond, and the complex numbers $ \alpha _{i}$ and $\alpha _{c}$ are the bonding coefficients satisfying $| \alpha _{i}| ^{2} +|\alpha _{c}| ^{2} =1$. In this manuscript, the AF model is generalized for the compound in which the chemical bonds are identical to the red one in Fig. 1 (a). To include the bonding correlation, the extended AF density matrix
\begin{eqnarray}
\rho _{ea} =  \left[
\begin{array}{c} 
\rho  \text{ \ \ }   \Delta   \\  \Delta  \text{ \ \ } \rho   
\end{array}
\right] 
\end{eqnarray}
is introduced to construct the correlation operators
\begin{eqnarray}
\begin{cases} 
d ^{(1)} = ( \sqrt{2} -1) [ \rho ^{(1)} \rho ^{(2)} ( I - \rho ^{(1)}) +   ( I - \rho ^{(1)}) \rho ^{(2)} \rho ^{(1)} ]   \\  
d ^{(2)} =  ( I - \rho ^{(1)})  \rho ^{(2)}  ( I - \rho ^{(1)}) \text{ \ \ \ \ \ \ \ \ \ \ \ \ \ \ \ \ \ \ \ \ \ \ \ \ \ \ \ \ }
\end{cases}
\end{eqnarray}
for the electron pair in the bonding region $\Omega$. Here $\rho$ and $\Delta$ are self-adjoint operators, $I$ denotes the identity operator, and the operators $\rho ^{(1)} = \rho + \Delta$ and $\rho ^{(2)} = \rho - \Delta$ can serve as the density matrices for the quasielectrons in Fig. 2 (a), which shows the 4-level dimer corresponding to the one-bond system. The correlation matrices $d ^{(1)}$ and $d ^{(2)}$ are denoted as the density-fluctuation and covalent-correlation operators because they represent the fluctuating charge and the correlation due to the covalent component, respectively. For convenience, the background is mentioned in section II, and the operators $d ^{(1)} $ and $d ^{(2)}$ are introduced in subsection III-A by considering the non-interacting and interacting parts of the one-bond Hamiltonian. The assumptions about the one-bond dimer are discussed in Appendix A, and an orbital transformation is mentioned in Appendix B to improve my model. The ionized and affinitive processes for the one-bond system are taken into account in subsection IV-A.

The quasielectron model for the binary compound is constructed in section III-B by considering the identical chemical bonds connecting the anions and cations. Figure 1 (b) shows such bonds in the zincblende-structure crystal as an example. The density-fluctuation and covalent-correlation operators are extended as $d ^{( \text{I} )}$ and $d ^{( \text{II} )}$ for the bonding correlation in the quasielectron system of the considered compound. I decompose such a system into the 4-level subsystems shown in Fig. 3 (a), and the energy difference to excite a carrier is obtained in section IV-B by considering the ionized/affinitive process.  When the compound forms an ideal crystal following the periodic boundary condition, each subsystem may correspond to a Bloch-type function. Near the ionic limit, as shown in Appendix C, my model can be supported by the coupled-cluster theory. On the other hand, we can see the importance of the bonding coefficients to the excited carrier near the covalent limit under the strong e-e repulsive strength, which is responsible for the Mott insulating behaviors in some AF systems \cite{Lee1,Liu}. The Hamiltonian family, which can correspond to the random Schr$\ddot{o}$dinger operators in the random-matrix theory \cite{Kirsc,Erdos,Lee2,Huang2}, is discussed in section V.  Actually we may generalize Eq. (2) to include a set of Hamiltonians by constructing the multiple-component quasielectrons. I note that the multiple-component functions can be used to introduce the vector bundles \cite{Bohm,Friedman,Banks} for the gauge theory. The compound system composed of different ionic-covalent dimers are discussed in Appendix D. The summary is made in section VI.

\section{Background}

The DDW Hamiltonian $H _{ddw} = \sum _{ {\bf k} \sigma } \chi _{ {\bf k} \sigma } ^\dag B _{ {\bf k} }  \chi _{ {\bf k} \sigma } $ \cite{Gerami} has been introduced for the AF quasielectrons with
\begin{eqnarray}
B _{ {\bf k} } =  \left[
\begin{array}{c} 
\varepsilon _{ {\bf k} } - \mu   \text{ \ \ \ \ \ \ \ }   \Delta _{ {\bf k} } \text{ \ \ \ \ }  \\  \text{ \ \ \ } \Delta _{ {\bf k} } ^{\ast} \text{ \ \ \ \ }  \varepsilon _{ {\bf k} + {\bf Q} } - \mu   
\end{array}
\right]
\end{eqnarray}
and $\chi _{ {\bf k} \sigma }  ^{\dag} = ( c _{ {\bf k} \text{ } \sigma } ^{\dag} , - i c _{ {\bf k}+{\bf Q} \text{ } \sigma } ^{\dag}  )$. Here $\mu$ denotes the chemical potential, the wave-vector {\bf k} belongs to the reduced Brillouin zone (RBZ),  ${\bf Q }$ equals the DDW ordering wavevector, $\sigma$ denotes the spin orientation $\uparrow$ or $\downarrow$, $\varepsilon _{ {\bf k} }$ and $ \varepsilon _{ {\bf k} + {\bf Q} }$ are the energy coefficients, $ \Delta _{ {\bf k} }$ is the DDW order parameter, and $ c _{ {\bf k} \text{ } \sigma }$ and $ c _{ {\bf k}+{\bf Q} \text{ } \sigma }$ are to annihilate electrons at $({\bf k} , \sigma )$ and $({\bf k} +{\bf Q}, \sigma )$, respectively. At half-filling, the zero-temperature chemical potential may lie in the gap between the two DDW energy bands, under which the gapless quasielectrons appear at the nodal points \cite{Gerami,Chakravarty}. The lower band is filled with the AF quasielectrons while the upper one is empty. Each filled eigenket $\varphi_{f}$ can be denoted by the two-component wavefunction
\begin{eqnarray}
| \varphi _{ f } \rangle =  \left[
\begin{array}{c} 
\varphi  ^{\prime} _{ f }  \\   \varphi ^{\prime \prime} _{ f } 
\end{array}
\right]
\end{eqnarray}
with the plane waves $\varphi  ^{\prime} _{ f } = a _{f} ^{\prime} exp (i {\bf k}_{f} \cdot {\bf r})$ and $\varphi ^{\prime \prime} _{ f }  = a _{f} ^{ \prime \prime}  exp (i ({\bf k}_{f} + {\bf Q} ) \cdot {\bf r})$. Here ${\bf k}_{f} \in $ RBZ, and the coefficients  $ a _{f} ^{\prime}$ and $ a _{f} ^{ \prime \prime}$ satisfying $( a _{f} ^{ \prime \ast }$, $ -i a _{f} ^{ \prime \prime \ast} ) B _{ {\bf k} } \propto ( a _{f} ^{ \prime \ast}$, $ -i a _{f} ^{ \prime \prime \ast} )$ are normalized such that $\langle \varphi _{f} | \varphi _{f} \rangle =1$. We may exchange the components in Eq. (5) to obtain 
\begin{eqnarray}
T | \varphi _{ f } \rangle =  \left[
\begin{array}{c} 
\varphi ^{\prime \prime} _{ f }  \\   \varphi  ^{\prime} _{ f }
\end{array}
\right],
\end{eqnarray}
which also represents the ket $\varphi _{ f }$, by the operator
\begin{eqnarray}
T =  \left[
\begin{array}{c} 
0 \text{ \ \ }  I  \\  I  \text{ \ \ }  0  
\end{array}
\right]. 
\end{eqnarray}
The extended AF density matrix  $ \sum  ( | \varphi _{ f } \rangle  \langle \varphi _{ f } | + T  | \varphi _{ f } \rangle  \langle \varphi _{ f } | T ^{\dag} ) $ is of the form of $\rho _{ea}$ given by Eq. (2) because it commutes with $T$. We can decompose $\rho _{ea}$ as 
\begin{eqnarray}
\rho _{ea}  = \rho ^{(1)} \otimes e _{1} e _{1} ^{\dag} + \rho ^{(2)} \otimes e _{2} e _{2} ^{\dag}. 
\end{eqnarray}
such that $\rho ^{(1)}  = \rho + \Delta = e _{1} ^{\dag} \rho _{ea} e _{1} $ and $\rho ^{(2)} = \rho - \Delta = e _{2} ^{\dag} \rho _{ea} e _{2} $, where $e _{1}  ^{\dag} = \frac{1}{ \sqrt{2} } (1,1) $ and $e _{2} ^{\dag} = \frac{1}{ \sqrt{2} } (1,-1) $.  The effective Hamiltonian $\hat{H_{D}}$ given by Eq. (49) in Ref. \cite{Huang1} can be obtained by generalizing $H _{ddw}$ for the extended AF density matrix. In addition to the orbital antiferromagnetism, the order parameter $\Delta _{ {\bf k} }$ is taken into account for d-wave superconductivity \cite{Chakravarty,Bena,Kee}. Moreover, we may use $\rho ^{(1)}$ and $\rho ^{(2)}$ to represent the quasielectrons in the one-bond system \cite{Huang1}. 

The 4-level dimer composed of the two spatial sites with up- and down-spin orientations has been introduced by considering the dimer Hamiltonian $H _{t-U} = \sum _{\sigma} (t _{AB} c _{A \sigma} ^{\dag} c _{B \sigma} +t _{AB} ^{*} c _{B \sigma} ^{\dag} c _{A \sigma}) + \hat{U}$ \cite{Matlak1}.  Here $A$ and $B$ represent the two spatial sites, $t _{AB}$ is the hopping coefficient,  the annihilators $c _{A \sigma}$ and $c _{B \sigma}$ follow  $\{ c _{B \sigma} ,c _{B \sigma ^{\prime} } \} = \{ c _{A \sigma} ,c _{A \sigma ^{\prime} } \} = \{ c _{A \sigma} ,c _{B \sigma ^{\prime} } \} = \{ c _{A \sigma} ,c _{B \sigma ^{\prime} } ^{\dag} \} =0$ and $\{ c _{A \sigma} ,c _{A \sigma ^{\prime} } ^{\dag} \} = \{ c _{B \sigma} ,c _{B \sigma ^{\prime} } ^{\dag} \} = \delta _{ \sigma \sigma ^{\prime} } $, the factor $\sum _{\sigma} (t _{AB} c _{A \sigma} ^{\dag} c _{B \sigma} +t _{AB} ^{*} c _{B \sigma} ^{\dag} c _{A \sigma})$ is responsible for the intra-bond hopping, and $\hat{U}$ denotes effective e-e interaction potential. To model the AF states, we shall note that such states occur as the up- and down-spin electrons repel each other when they are at the same sites. We can take
\begin{eqnarray}
\hat{U} = \int d ^{3} r ^{\prime} d ^{3} r ^{\prime \prime} U ( | {\bf r} ^{ \prime} - {\bf r} ^{ \prime \prime} | ) \psi _{ \uparrow } ^{\dag} ({\bf r} ^{\prime}) \psi _{ \uparrow } ( {\bf r} ^{\prime} ) \psi _{ \downarrow } ^{\dag} ({\bf r} ^{ \prime \prime}) \psi _{ \downarrow } ({\bf r} ^{ \prime \prime})
\end{eqnarray}
as the short-range repulsive potential to reduce the double occupancy \cite{Lee1} such that the electrons prefer $c _{A \uparrow} ^{\dag} c _{B \downarrow} ^{\dag} | 0 \rangle $ and $c _{B \uparrow} ^{\dag} c _{A \downarrow} ^{\dag} | 0 \rangle $ at half filling. Here $| 0 \rangle $ denotes the vacuum state,  the nonnegative function $U(|{\bf r}|)$ equals zero as $|{\bf r}|$ exceeds the short interacting length $a$, and $ \psi _{ \sigma } ( {\bf r} )$  is to annihilate the electron with the spin orientation $\sigma$ at position ${\bf r}$. Such a dimer can be used to model the ionic-covalent chemical bond, which separates atom $A$ from atom $B$ in Fig. 1 (a). Assume that the bonding electrons almost concentrate in $\Omega$ such that we can consider the approximation $ \langle {\bf r} | A \rangle = \langle {\bf r} | B \rangle =0$ as $ {\bf r} \notin \Omega$, where the $A$- and $B$-site orbitals $| A \rangle$ and $|B \rangle$ denote the normalized spatial parts of $c _{A \sigma} ^{\dag} | 0 \rangle$ and $c _{ B \sigma} ^{\dag} | 0 \rangle$, respectively. In this manuscript, I assume that $| A \rangle$ and $|B \rangle$ are localized near the corresponding atoms in the considered bond, as shown in Fig. 4. We can construct these site orbitals by considering the linear combinations of the atomic orbitals after truncating the tails outside the bonding region $\Omega$ and performing suitable orthogonalization to have the zero overlap integral  \cite{Levine}. Let $ \mathbb{C} ^{2} _{\Omega} $ be the vector space composed of the linear combinations of $| A \rangle$ and $|B \rangle$. The space $ \mathbb{C} ^{2} _{\Omega} $  is isomorphic to $ \mathbb{C} ^{2} $, and is a subspace of the Hilbert space $ L^{2} _{\Omega} $ composed of the square-integrable functions which equal zero outside $\Omega$. The Heitler-London correlated state $| \Psi _{c} \rangle = \frac{1}{ \sqrt{2} } ( c_{A \uparrow} ^{\dag} c_{B \downarrow} ^{\dag} +  c_{B \uparrow} ^{\dag} c_{A \downarrow} ^{\dag}) | 0 \rangle $, a linear combination of the AF states $c_{A \uparrow} ^{\dag} c_{B \downarrow} ^{\dag} | 0 \rangle $ and $c_{B \uparrow} ^{\dag} c_{A \downarrow} ^{\dag} | 0 \rangle $, has been taken into account in the covalent limit \cite{Fulde, Soos} when the 4-level bond is half-filled. To include the ionic part, we may consider the ionic Hubbard model \cite{Kampf, Buzatu} and modify $H _{t-U}$ as the one-bond Hamiltonian
\begin{eqnarray}
H _{b} = \sum _{\sigma} (t _{AB} c _{A \sigma} ^{\dag} c _{B \sigma} + t _{AB} ^{*} c _{B \sigma} ^{\dag} c _{A \sigma} + \varepsilon _{A}  c _{A \sigma} ^{\dag} c _{A \sigma} + \varepsilon _{B}  c _{B \sigma} ^{\dag} c _{B \sigma}) + \hat{U},
\end{eqnarray}
which is suitable to model the polar molecule \cite{Prasad}, to introduce the on-site energies $\varepsilon _{A}$ and $\varepsilon _{B}$. In the above equation, $\sum _{\sigma} (t _{AB} c _{A \sigma} ^{\dag} c _{B \sigma} + t _{AB} ^{*} c _{B \sigma} ^{\dag} c _{A \sigma} + \varepsilon _{A}  c _{A \sigma} ^{\dag} c _{A \sigma} + \varepsilon _{B}  c _{B \sigma} ^{\dag} c _{B \sigma}) = H _{b} - \hat{U} $ is the non-interacting part of $H _{b}$ while the e-e potential $\hat{U}$ serves as the interacting part. In the following, assume that $\varepsilon _{B} > \varepsilon _{A}$ such that atom $A$ has the higher electronegativity. The two-electron wavefunction becomes the uncorrelated state $ | \Psi _{i} \rangle = c _{A \uparrow} ^{\dag} c _{A \downarrow} ^{\dag} | 0 \rangle$ in the ionic limit, so we shall take both $ | \Psi _{i} \rangle$ and $ | \Psi _{c} \rangle$ in general and consider the bonding wavefunction given by Eq. (1) at half filling. The assumptions about Eq. (10) are discussed in Appendix A. While there exists another state $ | \Psi _{BB} \rangle = c ^{\dagger} _{B \uparrow} c ^{\dagger} _{B \downarrow} |0 \rangle$, as mentioned in Appendix B, its contribution is small and we can include it just by performing an orbital transformation. For convenience, I introduce my model without considering $ | \Psi _{BB} \rangle$ in the main text, and include its contribution in Appendix B by using such a transformation to preserve the form of Eq. (1).      

Because the bonding electrons in the covalent limit are described by the linear combination of the AF states $ c_{A \uparrow} ^{\dag} c_{B \downarrow} ^{\dag} | 0 \rangle$ and $ c_{B \uparrow} ^{\dag} c_{A \downarrow} ^{\dag} | 0 \rangle$, it is natural to try the extended AF density matrix $\rho _{ea}$ to construct the quasielectron orbitals of the one-bond system in Fig. 1 (a). In each AF state one electron is located at $ | A \rangle$ while the other one is located at $ |B \rangle$, so we shall take the matrices $\rho ^{(1)}$ and $\rho ^{(2)}$ in Eq. (8) as $| A \rangle \langle A |$ and $| B \rangle \langle B |$ in the covalent limit. On the other hand, both electrons occupy $ | A \rangle$ in the ionic limit and thus these two matrices should equal $| A \rangle \langle A |$ as $| \Psi _{i} \rangle$ is dominated. By taking \cite{Huang1}
\begin{eqnarray}
\rho ^{(1)} = | A \rangle \langle A | \text{ and } \rho ^{(2)} = | L \rangle \langle L |              
\end{eqnarray}       
with $ | L \rangle = \alpha _ {i} | A \rangle + \alpha _{c} | B \rangle$ in $\mathbb{C} ^{2} _{ \Omega }$, the matrix $\rho ^{(1)}$ just represents the quasielectron occupying $| A \rangle$ under both limits. In addition, the matrix $\rho ^{(2)}$ corresponds to the other one jumping to $| A \rangle$ from $| B \rangle$ as $| \Psi _{i} \rangle$ becomes significant. The operator $\rho _{sb} = \frac{1}{2} (\rho ^{(1)} + \rho ^{(2)} + d ^{(1)} )$ satisfies $\langle {\bf r} ^{\prime} | \rho _{sb} | {\bf r} ^{ \prime \prime } \rangle = \langle \Psi _{b} | \psi _{ \uparrow } ^{ \dagger } ({\bf r} ^{ \prime } ) \psi _{ \uparrow } ({\bf r} ^{\prime \prime} ) | \Psi _{b} \rangle =  \langle \Psi _{b} | \psi _{ \downarrow } ^{ \dagger } ({\bf r} ^{ \prime } ) \psi _{ \downarrow } ({\bf r} ^{\prime \prime} ) | \Psi _{b} \rangle  $ and thus serves as the one-electron density matrix, which is spin-degenerate because $\langle \Psi _{b} | \psi _{ \sigma } ^{ \dagger } ({\bf r} ^{\prime}) \psi _{ - \sigma } ({\bf r}^{\prime \prime} ) | \Psi _{b} \rangle=0$ for $\sigma = \uparrow$ and $\downarrow$.  

\section{Quasielectrons in ionic-covalent bonds}

In this section, the energy forms are constructed for the quasielectrons in the half-filled bonding systems. For convenience, I discuss the one-bond system based on the 4-level dimer \cite{Matlak1} in subsection III-A, and extend the results to the compound system composed of identical dimers \cite{Matlak2} in subsection III-B. 

\section*{III-A One-bond system}

Consider the two uncorrelated states $| \Phi _{b} \rangle = c _{A \uparrow} ^{\dag} c _{ L \downarrow} ^{\dag} | 0 \rangle = \alpha _{i} | \Psi _{i} \rangle + \alpha _{c} c _{ A \uparrow} ^{\dag} c _{B \downarrow} ^{\dag} | 0 \rangle  $  and  $| \Phi _{b} ^{\prime} \rangle = c _{L \uparrow} ^{\dag} c _{ A \downarrow} ^{\dag} | 0 \rangle = \alpha _{i} | \Psi _{i} \rangle + \alpha _{c} c _{ B \uparrow} ^{\dag} c _{A \downarrow} ^{\dag} | 0 \rangle  $, which can be obtained from  $| \Psi _{b} \rangle$ by substituting the AF states $ c _{ A \uparrow} ^{\dag} c _{B \downarrow} ^{\dag} | 0 \rangle  $ and $c _{ B \uparrow} ^{\dag} c _{A \downarrow} ^{\dag} | 0 \rangle$, respectively, for the covalent component $| \Psi _{c} \rangle$ in Eq. (1). Here $c _{L \sigma}$ is the operator to annihilate the electron with the spin orientation $\sigma$ at $| L \rangle$ in the bonding region $\Omega$ in Fig. 1 (a). The matrices $\rho ^{(1)}$ and $\rho ^{(2)}$ in Eq. (11) correspond to the up- and down-spin electrons in $| \Phi _{b} \rangle$, and correspond to the down- and up-spin ones in $ | \Phi _{b} ^{\prime} \rangle$. So $\rho ^{(1)}$ and $\rho ^{(2)}$ are for the quasielectrons with the opposite spin orientations in the one-bond system, and can represent the occupied levels of the half-filled 4-level dimer in Fig. 2 (a) if $| 1 \rangle \otimes | \sigma \rangle = |A \rangle \otimes | \sigma \rangle $ and $| 2 \rangle \otimes | - \sigma \rangle =  | L \rangle \otimes | - \sigma \rangle $. The two unoccupied levels $| \bar{1} \rangle \otimes | \sigma \rangle$ and $| \bar{2} \rangle \otimes | - \sigma \rangle$ in Fig. 2(a) serve as $ | B \rangle \otimes | \sigma \rangle$ and $| \bar{L} \rangle \otimes | - \sigma \rangle$ in such a one-bond system, where $| \bar{L} \rangle = \alpha _{c} ^{*} | A \rangle - \alpha _{i} ^{*} | B \rangle \bot | L \rangle$ is a ket in $\mathbb{C} ^{2} _{\Omega}$. The uncorrelated energy  $\langle \Phi _{b} | H _{b} | \Phi _{b}  \rangle = \langle \Phi ^{\prime} _{b} | H _{b} | \Phi _{b} ^{\prime}  \rangle$ equals
\begin{eqnarray}
 tr  (\rho ^{(1)} + \rho ^{(2)}) H_{sb} + \int _{ {\bf r} ^{ \prime },{\bf r} ^{ \prime \prime} \in \Omega} d ^{3} r ^{\prime} d ^{3} r ^{ \prime \prime } U ( | {\bf r} ^{\prime} - {\bf r} ^{\prime \prime} | ) \langle {\bf r} ^{\prime} | \rho ^{(1)} | {\bf r} ^{\prime} \rangle \langle {\bf r} ^{\prime \prime} | \rho ^{(2)} | {\bf r} ^{\prime \prime} \rangle,
\end{eqnarray}
where $H_{sb} = t _{AB}  | A \rangle \langle B | + t _{AB} ^{*} | B \rangle \langle A | + \varepsilon _{A}  |  A \rangle \langle A | + \varepsilon _{B} | B  \rangle  \langle B |$ results from the non-interacting part of $H_{b}$. The factors $t _{AB}  | B \rangle \langle A | + t _{AB} ^{*} | A \rangle \langle B |$ and $\varepsilon _{A}  |  A \rangle \langle A | + \varepsilon _{B} | B  \rangle  \langle B |$ of $H _{sb}$ are responsible for the intra-bond hopping and on-site energy difference, respectively. 

While we may construct the qausielectron density matrices by Eq. (11), the wavefunction $| \Psi _{b} \rangle$ in Eq. (1) is different from the uncorrelated functions $| \Phi _{b} \rangle$ and $| \Phi _{b} ^{\prime} \rangle$ when $ \alpha _{c} \neq 0$. To obtain the bonding energy $E _{b} = \langle \Psi _{b} | H _{b} | \Psi _{b}  \rangle$, therefore, we shall include the correlation contributions corresponding to the blue dash curve in Fig. 2 (a) by calculating the difference $\langle \Psi _{b} | H _{b} | \Psi _{b}  \rangle - \langle \Phi _{b} | H _{b} | \Phi _{b}  \rangle = \langle \Psi _{b} | H _{b} | \Psi _{b}  \rangle -  \langle \Phi ^{\prime} _{b} | H _{b} | \Phi _{b} ^{\prime}  \rangle$. The non-interacting part of $H_{b}$ induces the correlation energy $ tr H_{sb} d ^{(1)} = \langle \Psi _{b} | (H _{b} - \hat{U}) | \Psi _{b}  \rangle - \langle \Phi _{b} | (H _{b} - \hat{U}) | \Phi _{b}  \rangle = \langle \Psi _{b} | (H _{b} - \hat{U}) | \Psi _{b}  \rangle - \langle \Phi ^{\prime} _{b} | (H _{b} - \hat{U}) | \Phi ^{\prime} _{b}  \rangle $. We can take $I = | A \rangle \langle A| +  | B \rangle \langle B | $, which is the identity operator on $\mathbb{C} ^{2} _{\Omega}$, in Eq. (3) to introduce the density-fluctuation operator $ d ^{(1)} $. The operator $d ^{(1)}$ results from the factor $\alpha _{i} \alpha _{c} ^{*}  \langle \Psi _{c} | (H _{b} - \hat{U}) | \Psi _{i}  \rangle + \alpha _{c} \alpha _{i} ^{*}  \langle \Psi _{i} | (H _{b} - \hat{U}) | \Psi _{c}  \rangle  $ when we calculate $\langle \Psi _{b} | (H _{b} - \hat{U}) | \Psi _{b}  \rangle $, and becomes zero if the ionic and covalent components do not coexist. The up- or down-spin density $\langle {\bf r} | \rho _{sb} | {\bf r} \rangle $ at position ${\bf r}$ includes the factor $\langle {\bf r} | d ^{(1)} | {\bf r} \rangle$, which follows $\int _{ {\bf r} \in \Omega}  d ^{3} r  \langle {\bf r} | d ^{(1)} | {\bf r} \rangle=tr d ^{(1)} =  ( \sqrt{2} -1) tr [ ( I - \rho ^{(1)}) \rho ^{(1)} \rho ^{(2)}  +  \rho ^{(2)} \rho ^{(1)}  ( I - \rho ^{(1)}) ]=0 $ because $( I - \rho ^{(1)}) \rho ^{(1)}= \rho ^{(1)}  ( I - \rho ^{(1)})=0$ under Eq. (11). Therefore, $\langle {\bf r} | d ^{(1)} | {\bf r} \rangle$ has no contribution to the total charge $2 \int _{ {\bf r} \in \Omega}  d ^{3} r \langle {\bf r} | \rho _{sb} | {\bf r} \rangle $, and represents the fluctuating charge density due to the coexistence of the ionic and covalent components. I note that $\delta \rho$, the deviation from the average density \cite{Zhang} due to the correlation in the quantum Hall effect \cite{DHLee,You}, also has no contribution to the total charge. 

To include the correlation energy due to $\hat{U}$ in the quasielectron space, we can further introduce $d ^{(2)}$ to rewrite  $ \langle \Psi _{b} | \hat{U} | \Psi _{b} \rangle - \langle \Phi _{b} | \hat{U} | \Phi _{b} \rangle$ or $\langle \Psi _{b} | \hat{U} | \Psi _{b} \rangle - \langle \Phi ^{\prime} _{b} | \hat{U} | \Phi _{b} ^{\prime} \rangle$ as $\int _{ {\bf r} ^{\prime},{\bf r} ^{\prime \prime} \in \Omega} d ^{3} r ^{\prime} d ^{3} r ^{\prime \prime} U ( | {\bf r} ^{\prime} - {\bf r} ^{\prime \prime} | ) [ \langle {\bf r} ^{\prime} | \rho ^{(1)} | {\bf r} ^{\prime \prime } \rangle \langle {\bf r} ^{\prime \prime} | d ^{(2)} | {\bf r} ^{ \prime } \rangle + \langle {\bf r} ^{\prime} | \rho ^{(1)} | {\bf r} ^{\prime } \rangle \langle {\bf r} ^{\prime \prime} | d ^{(1)} | {\bf r} ^{\prime \prime } \rangle] $. We can take $I= |A\rangle \langle A| + | B \rangle \langle B| $ in Eq. (3) to introduce  $d ^{(2)}$ on $\mathbb{C} ^{2} _{ \Omega} $. The operator $d ^{(2)}$ comes from the factor $ | \alpha _{c} | ^{2} \langle \Psi _{c} | \hat{U} | \Psi _{c} \rangle$ when we calculate $\langle \Psi _{b} | \hat{U} | \Psi _{b} \rangle$, and becomes zero when  $\alpha _{c} =0$. Hence $d ^{(2)}$ represents the covalent correlation, and we shall include both $d ^{(1)}$ and $d ^{(2)}$ for the blue dash curve in Fig. 2 (a). By including the correlation contributions, we have 
\begin{eqnarray}
E_{b} =  tr (\rho ^{(1)} + \rho ^{(2)}) H_{sb} + \int _{ {\bf r} ^{ \prime },{\bf r} ^{ \prime \prime } \in \Omega} d ^{3} r ^{ \prime } d ^{3} r ^{\prime \prime} U ( | {\bf r} ^{\prime} - {\bf r} ^{\prime \prime} | ) \langle {\bf r} ^{\prime} | \rho ^{(1)} | {\bf r} ^{\prime} \rangle \langle {\bf r} ^{\prime \prime} | \rho ^{(2)} | {\bf r} ^{\prime \prime} \rangle +
\end{eqnarray}
\[
 tr H_{sb} d ^{(1)} + \int _{ {\bf r} ^{\prime},{\bf r} ^{\prime \prime} \in \Omega} d ^{3} r ^{\prime} d ^{3} r ^{\prime \prime} U ( | {\bf r} ^{\prime} - {\bf r} ^{\prime \prime} | ) \langle {\bf r} ^{\prime} | \rho ^{(1)} | {\bf r} ^{\prime} \rangle \langle {\bf r} ^{\prime \prime} | d ^{(1)} | {\bf r} ^{\prime \prime} \rangle + \text{ \ \ \ \ \ }
\]
\[
\int _{ {\bf r} ^{ \prime },{\bf r} ^{\prime \prime} \in \Omega} d ^{3} r ^{\prime} d ^{3} r ^{\prime \prime} U ( | {\bf r} ^{\prime} - {\bf r} ^{\prime \prime} | ) \langle {\bf r} ^{\prime} | \rho ^{(1)} | {\bf r} ^{\prime \prime} \rangle \langle {\bf r} ^{\prime \prime} | d ^{(2)} | {\bf r} ^{\prime} \rangle \text{ \ \ \ \ \ \ \ \ \ \ \ \ \ \ \ \ \ \ \ \ \ \ \ \ \ }
\]
from Eq. (12) when the ionic-covalent chemical bond in Fig. 1 (a) is half-filled. In the above equation, the factors in the second and third lines are due to the bonding correlation. The above equation provides the energy form of the density matrix $\rho  _{ea}$, which can be decomposed into $\rho ^{(1)} = e _{1} ^{\dag} \rho  _{ea} e _{1}$ and  $\rho ^{(2)} = e _{2} ^{\dag} \rho  _{ea} e _{2}$, for the one-bond system.

\section*{III-B Compound system}

Consider the binary compound where the anions and cations are connected by the identical ionic-covalent bonds, and assume that all the bonds are well-separated without overlap.  Figure 1 (b) shows such bonds in the zincblende-structure crystal \cite{Grosso}, in which each atom provides 4 site orbitals, as an example. For convenience, I parameterize these bonds by the integer parameter $j=1 \sim N$ and denote the bonding region of the $j$-th bond as $\Omega _{j}$, where $N$ is the total number of the bonds. Each bond in the compound is a 4-level dimer just as the one-bond system in Fig. 1 (a). Assume that there exists the one-to-one mapping ${\bf R} _{j}$ to relate any position ${\bf r} _{j} \in \Omega _{j}$ in the $j$-th bond to ${\bf r} \in \Omega$ in Fig. 1 (a) by ${\bf R} _{j} ({\bf r}) = {\bf r} _{j}$ and ${\bf R} _{j} ^{-1} ({\bf r} _{j}) = {\bf r}$ such that the distance $|{\bf r}_{1}-{\bf r} _{2}|$ between ${\bf r}_{1}$ and ${\bf r}_{2} \in \Omega$ equals $|{\bf R} _{j} ( {\bf r} _{1} )- {\bf R} _{j}({\bf r} _{2})| $ for all $j$. Therefore, every ionic-covalent chemical bond in the compound is identical to that in Fig. 1 (a). Let $| A, j \rangle $ and $| B, j \rangle $ as the kets mapped from $| A \rangle $ and $| B \rangle $ under ${\bf R}_{j}$, respectively. The space $\mathbb{C} ^{2} _{ \Omega _{j} }$ spanned by $| A, j \rangle $ and $| B, j \rangle $ is a subspace of the Hilbert space $L^{2} _{ \Omega _{j} }$ composed of square-integrable functions which equal zero outside $\Omega _{j}$, and any two kets in $L^{2} _{ \Omega _{l} }$ and $L^{2} _{ \Omega _{p} }$ are orthogonal to each other when $l \neq p$. We can introduce the space $\mathbb{C} ^{N} \otimes \mathbb{C} ^{2} _{\Omega} = \mathbb{C} ^{2} _{\Omega _{1} }  \oplus \mathbb{C} ^{2} _{\Omega _{2} }  \oplus \textellipsis \oplus \mathbb{C} ^{2} _{\Omega _{N} }$ for the compound, and choose a set of orthonormal basis $\{  | w _{j} \rangle \}$ in $\mathbb{C} ^{N}$ to represent $| A, j \rangle $ and $| B, j \rangle $  by $|w _{j} \rangle \otimes | A \rangle $ and $|w _{j} \rangle \otimes | B \rangle $, respectively. The space $\mathbb{C} ^{N} \otimes \mathbb{C} ^{2} _{\Omega}$ is a subspace of $\mathbb{C} ^{N} \otimes L ^{2} _{\Omega} = L ^{2} _{\Omega _{1} }  \oplus L ^{2} _{\Omega _{2} }  \oplus \textellipsis \oplus L ^{2} _{\Omega _{N} }$, and the position ket $| {\bf r} _{j} \rangle$ corresponding to the position ${\bf r} _{j} \in \Omega _{j} $ is taken as $|w _{j} \rangle \otimes |  {\bf R} _{j} ^{-1} ({\bf r} _{j})  \rangle $ to perform the integral in $\mathbb{C} ^{N} \otimes L ^{2} _{\Omega}$. For any function $F _{\Omega}$ defined on $\Omega$ and the position ${\bf r} _{j _{1}} \in {\Omega}_{j _{1}} $, we have $(\langle {\bf r} _{j _{1} } |)(| w _{ j _{2} } \rangle \otimes | F _{\Omega} \rangle) = ( \langle w _{ j _{1} } | \otimes \langle {\bf R} _{ j _{1} } ^{ -1 } ( {\bf r} _{ j _{1} } ) | )( | w _{ j _{2} } \rangle \otimes | F _{\Omega} \rangle ) = \delta _{ j _{1} , j _{2} } \langle  {\bf R} _{ j _{1} } ^{ -1 } ( {\bf r} _{ j _{1} } ) | F _{\Omega} \rangle =  \delta _{ j _{1} , j _{2} } F _{\Omega} ({\bf R} _{ j _{1} } ^{ -1 } ( {\bf r} _{ j _{1} } ) )$.

By mapping the one-bond system in Fig. 1 (a) to the identical bonds in the compound, at half filling we can transfer $ \rho ^{(1)}$ and $ \rho ^{(2)}$ in Eq. (11) to the $j$-th bond and obtain $ \rho _{ w _{j} } ^{(1)} = | w _{j} \rangle \langle w _{j} | \otimes |A \rangle \langle A |$ and $\rho _{ w _{j} } ^{(2)} = | w _{j} \rangle \langle w _{j} | \otimes |L \rangle \langle L|$ for the quasielectrons at $ | A, j \rangle$ and $ | L, j \rangle $. Here $ | L, j \rangle = \alpha _{i} | A, j \rangle +  \alpha _{c} | B, j \rangle$. The matrices
\begin{eqnarray}
\rho ^{( \text{I} )} = \sum _{j=1} ^{N} \rho _{ w _{j} } ^{(1)} \text{ \ \ and \ \ } \rho ^{( \text{II} )} = \sum _{j=1} ^{N} \rho _{ w _{j} } ^{(2)}
\end{eqnarray}
for the half-filled compound are of the opposite spin orientations just as $ \rho ^{(1)}$ and $ \rho ^{(2)}$ in the one-bond system, and the corresponding extended AF density matrix is $\rho ^{( \text{I} )} e _{1} e _{1} ^{\dag} + \rho ^{( \text{II} )} e _{2} e _{2} ^{\dag}$. In fact, we can take $I _{ \mathbb{C} ^{N} } $ as the identity operator on $\mathbb{C} ^{N}$ and rewrite Eq. (14) by $\rho ^{( \text{I} )} = I _{ \mathbb{C} ^{N} } \otimes \rho ^{(1)} $ and $\rho ^{( \text{II} )} = I _{ \mathbb{C} ^{N} } \otimes \rho ^{(2)} $ to see that $\rho ^{( \text{I} )}$ and $\rho ^{( \text{II} )}$ are the natural extensions of $ \rho ^{(1)}$ and $ \rho ^{(2)}$, respectively. To include the bonding correlation due to the $j$-th bond, we shall substitute $ d _{ w _{j} } ^{(1)} = | w _{j} \rangle \langle w _{j} | \otimes d ^{(1)}$ and $ d _{ w _{j} } ^{(2)} = | w _{j} \rangle \langle w _{j} | \otimes d ^{(2)}$ for the operators in Eq. (3). The correlation operators $d ^{( \text{I} )}=  \sum _{j} d _{ w _{j} } ^{(1)} = I _{ \mathbb{C} ^{N} } \otimes d ^{(1)}$ and $d ^{( \text{II} )}= \sum _{j} d _{ w _{j} } ^{(2)} =  I _{ \mathbb{C} ^{N} } \otimes d ^{(2)}$ satisfy
\begin{eqnarray}
\begin{cases} 
d ^{( \text{I} )} = ( \sqrt{2} -1) [ \rho ^{( \text{I} )} \rho ^{( \text{II} )} ( I - \rho ^{( \text{I} )}) +   ( I - \rho ^{( \text{I} )}) \rho ^{( \text{II} )} \rho ^{( \text{I} )} ] \\  
d ^{( \text{II} )} = ( I - \rho ^{( \text{I} )})  \rho ^{( \text{II} )}  ( I - \rho ^{( \text{I} )})   \text{ \ \ \ \ \ \ \ \ \ \ \ \ \ \ \ \ \ \ \ \ \ \ \ \ \ \ \ \ }
\end{cases},
\end{eqnarray}
and the matrix $\rho _{sC} = \frac{1}{2}(\rho ^{( \text{I} )}  + \rho ^{( \text{II} )} +d ^{( \text{I} )} ) = I _{ \mathbb{C} ^{N} } \otimes \rho _{sb}$ yields the density $Q({\bf r} _{j} ) = 2 \langle {\bf r} _{j} | \rho _{sC} | {\bf r} _{j} \rangle = 2 \langle {\bf R} _{j} ^{-1} ({\bf r} _{j}) | \rho _{sb} |  {\bf R} _{j} ^{-1} ({\bf r} _{j}) \rangle $ at ${\bf r} _{j} \in \Omega _{j}$. In the above equation, the matrix $I$ denotes the identity operator on $\mathbb{C} ^{N} \otimes \mathbb{C} ^{2} _{\Omega} $, and we can interpret $d ^{( \text{I} )}$ and $d ^{( \text{II} )}$ as the density-fluctuation and covalent-correlation operators of the quasielectron system in the compound because they serve as $d ^{(1)}$ and $d ^{(2)}$. For the ideal crystal, we may discuss the correlation under the crystal symmetry imposed on $ \rho ^{( \text{I} )}$ and $ \rho ^{( \text{II} )}$ based on Eq. (15).  

When the distances between different bonds are larger than $a$, the interacting length of $\hat{U}$, there is no inter-bond e-e interaction in the half-filled compound. Therefore, the two quasielectrons in a specific bond only interact with each other just as those in the one-bond system in Fig. 1 (a), and the e-e energy term is composed of  
\[
\text{(i) \ \ } \sum _{j=1} ^{N} \int _{ {\bf r} ^{\prime} _{ j },{\bf r} ^{\prime \prime} _{ j } \in \Omega _{j} } d ^{3} r ^{\prime} _{ j } d ^{3} r ^{\prime \prime} _{ j } U ( | {\bf r} ^{\prime} _{ j } -  {\bf r} ^{\prime \prime} _{ j } | ) \langle {\bf r} ^{\prime} _{ j } | \rho ^{( \text{I} )} | {\bf r} ^{\prime} _{ j } \rangle \langle {\bf r} ^{\prime \prime} _{ j }  | \rho ^{( \text{II} )} | {\bf r} ^{\prime \prime} _{ j }  \rangle 
\]
\begin{eqnarray}
\text{(ii) \ }  \sum _{j=1} ^{N} \int _{ {\bf r} ^{\prime} _{ j },{\bf r} ^{\prime \prime} _{ j } \in \Omega _{j} } d ^{3} r ^{\prime} _{ j } d ^{3} r ^{\prime \prime} _{ j } U ( | {\bf r} ^{\prime} _{ j } - {\bf r} ^{\prime \prime} _{ j } | ) \langle {\bf r} ^{\prime} _{ j } | \rho ^{( \text{I} )} | {\bf r} ^{\prime} _{ j } \rangle \langle {\bf r} ^{\prime \prime} _{ j }  | d ^{( \text{I} )} |{\bf r} ^{\prime \prime} _{ j }  \rangle 
\end{eqnarray}
\[
\text{(iii) } \sum _{j=1} ^{N} \int _{ {\bf r} ^{\prime} _{ j },{\bf r} ^{\prime \prime} _{ j } \in \Omega _{j} } d ^{3} r ^{\prime} _{ j } d ^{3} r ^{\prime \prime} _{ j } U ( | {\bf r} ^{\prime} _{ j } -  {\bf r} ^{\prime \prime} _{ j } | ) \langle {\bf r} ^{\prime} _{ j } | \rho ^{( \text{I} )} | {\bf r} ^{\prime \prime } _{ j } \rangle \langle {\bf r} ^{\prime \prime} _{ j }  | d ^{( \text{II} )} | {\bf r} ^{\prime} _{ j }  \rangle. \text{ \ }
\]
Equation (16)-(i) corresponds to the second term in Eq. (12) and can be obtained without considering the bonding correlation. On the other hand, Eqs. (16)-(ii) and (16)-(iii) correspond to the last two terms in Eq. (13) and provide the correlation contributions.  

To include the energy resulting from the non-interacting term $H _{sb}$, we shall consider the factor $2 tr \rho _{sC} ( \sum _{j=1} ^{N} | w _{j} \rangle \langle w _{j} | \otimes H_{sb} )  =  2 tr \rho _{sb} H _{sb} \times N$ in the compound system. In addition to the intra-bond hopping in $H _{sb}$, the inter-bond hopping 
\begin{eqnarray}
H _{hop} =  \sum _{ j \neq j ^{\prime} } t _{ j \xi , j ^{\prime} \xi ^{\prime} } | w _{ j } \rangle \langle w _{j ^{\prime} } | \otimes | \xi \rangle \langle \xi ^{\prime} |
\end{eqnarray}
should be taken into account to relate different chemical bonds  \cite{Matlak2}. Here  $ | \xi \rangle \text{ and }  | \xi ^{\prime}  \rangle \in \{ | A \rangle , |B \rangle \} $, and each coefficient $t _{ j \xi , j ^{\prime} \xi ^{\prime} } = t^{*} _{j ^{\prime} \xi ^{\prime}, j \xi } $ is for the jump from $| w _{ j ^{\prime} } \rangle \otimes | \xi ^{\prime} \rangle$ to $| w _{  j } \rangle \otimes | \xi \rangle$. Therefore, we shall introduce the non-interacting Hamiltonian $H _{sC} = H _{hop}+ \sum _{j=1} ^{N} | w _{j} \rangle \langle w _{j} | \otimes H_{sb} $ and include the energy $2 tr \rho _{sC} H _{sC}$. In this manuscript I consider the short-range hopping, so $t _{ j \xi , j ^{\prime} \xi ^{\prime} }=0$ if the distance between the $j$-th and $j^{\prime}$-th bonds is longer than a specific length. The energy for the half-filled compound system is 
\[
E_{Cr} = tr ( \rho ^{ ( \text{I} ) } + \rho ^{ ( \text{II} ) } ) H _{sC} + \sum _{j=1} ^{N} \int _{ {\bf r} ^{\prime} _{ j },{\bf r} ^{\prime \prime} _{ j } \in \Omega _{j} } d ^{3} r ^{\prime} _{ j } d ^{3} r ^{\prime \prime} _{ j } U ( | {\bf r} ^{\prime} _{ j } -  {\bf r} ^{\prime \prime} _{ j } | ) \langle {\bf r} ^{\prime} _{ j } | \rho ^{( \text{I} )} | {\bf r} ^{\prime} _{ j } \rangle  \langle {\bf r} ^{\prime \prime} _{ j }  | \rho ^{( \text{II} )} | {\bf r} ^{\prime \prime} _{ j }  \rangle +
\]
\begin{eqnarray}
trH _{sC} d^{ ( \text{I} )} + \sum _{j=1} ^{N} \int _{ {\bf r} ^{\prime} _{ j },{\bf r} ^{\prime \prime} _{ j } \in \Omega _{j} } d ^{3} r ^{\prime} _{ j } d ^{3} r ^{\prime \prime} _{ j } U ( | {\bf r} ^{\prime} _{ j } - {\bf r} ^{\prime \prime} _{ j } | ) \langle {\bf r} ^{\prime} _{ j } | \rho ^{( \text{I} )} | {\bf r} ^{\prime} _{ j } \rangle \langle {\bf r} ^{\prime \prime} _{ j }  | d ^{( \text{I} )} |{\bf r} ^{\prime \prime} _{ j }  \rangle  +
\end{eqnarray}
\[
 \sum _{j=1} ^{N} \int _{ {\bf r} ^{\prime} _{ j },{\bf r} ^{\prime \prime} _{ j } \in \Omega _{j} } d ^{3} r ^{\prime} _{ j } d ^{3} r ^{\prime \prime} _{ j } U ( | {\bf r} ^{\prime} _{ j } -  {\bf r} ^{\prime \prime} _{ j } | ) \langle {\bf r} ^{\prime} _{ j } | \rho ^{( \text{I} )} | {\bf r} ^{\prime \prime } _{ j } \rangle \langle {\bf r} ^{\prime \prime} _{ j }  | d ^{( \text{II} )} | {\bf r} ^{\prime} _{ j }  \rangle. \text{ \ \ \ \ \ \ \ \ \ \ \ \ \ \ \ \ \ } 
\]

In addition to $\{ | w _{j} \rangle \}$, we can choose another orthonormal complete set $ \{ |  \eta _{j} \rangle \}$ in $ \mathbb{C} ^{N}$ to describe the quasielectron system of the compound. For an example, it is important to choose the set composed of the Bloch-type functions in $\mathbb{C} ^{N}$ when the considered compound is an ideal crystal following the periodic boundary condition. The matrices in Eqs. (14) and (15) can be rewritten as $\rho ^{( \text{I} )} = \sum _{j=1} ^{N} \rho _{ \eta _{j} } ^{(1)}$, $\rho ^{( \text{II} )} = \sum _{j=1} ^{N} \rho _{ \eta _{j} } ^{(2)}$, $d ^{( \text{I} )} = \sum _{j=1} ^{N} d_{ \eta _{j} } ^{(1)}$, and $d ^{( \text{II} )} = \sum _{j=1} ^{N} d_{ \eta _{j} } ^{(2)}$, where $\rho _{ \eta _{j} } ^{(1)} = | \eta _{j} \rangle \langle \eta _{j} | \otimes | A \rangle \langle A |$, $\rho _{ \eta _{j} } ^{(2)} = | \eta _{j} \rangle \langle \eta _{j} | \otimes | L \rangle \langle L |$, $d _{ \eta _{j} } ^{(1)} = | \eta _{j} \rangle \langle \eta _{j} | \otimes d ^{(1)}$, and $d _{ \eta _{j} } ^{(2)} = | \eta _{j} \rangle \langle \eta _{j} | \otimes d ^{(2)}$. Every $| \eta _{j} \rangle$ can correspond to the 4-level dimer in Fig. 2 (a) if we take $\rho _{ \eta _{j} } ^{(1)} $ and $\rho _{ \eta _{j} } ^{(2)}$ as the two quaielectrons at $| 1\rangle = | \eta _{j} \rangle \otimes | A \rangle $ and $| 2 \rangle = | \eta _{j} \rangle \otimes | L \rangle $. The spatial parts of the two empty orbitals are $| \bar{1} \rangle = | \eta _{j} \rangle \otimes |B \rangle $ and $ | \bar{2} \rangle = | \eta _{j} \rangle \otimes | \bar{L} \rangle $, respectively. The operators $d _{ \eta _{j} } ^{(1)}$ and $d _{ \eta _{j} } ^{(2)}$ are determined by $\rho _{ \eta _{j} } ^{(1)} $ and $\rho _{ \eta _{j} } ^{(2)}$ because $d _{ \eta _{j} } ^{(1)} = ( \sqrt{2} -1) [ \rho _{ \eta _{j} } ^{(1)} \rho _{ \eta _{j} } ^{(2)} ( I - \rho _{ \eta _{j} } ^{(1)}) +   ( I - \rho _{ \eta _{j} } ^{(1)}) \rho _{ \eta _{j} } ^{(2)} \rho _{ \eta _{j} } ^{(1)} ]$ and $d _{ \eta _{j} } ^{(2)} = ( I- \rho _{ \eta _{j} } ^{(1)} )\rho _{ \eta _{j} } ^{(2)} ( I - \rho _{ \eta _{j} } ^{(1)}) $, and we can interpret $d _{ \eta _{j} } ^{(1)}$ and $d _{ \eta _{j} } ^{(2)}$ as the correlation contributions of the two quasielectrons in subsystem $\eta _{j}$. Therefore, the half-filled quasielectron system to model the compound are decomposed into the 4-level subsystems as shown in Fig. 3 (a), where the blue dash curves denote the corresponding correlation contributions. In addition, the two quasielectrons in each subsystem $\eta _{j}$ are correlated just as those described by $\rho ^{(1)}$ and $\rho ^{(2)}$ in the one-bond system.

\section{Electron affinity and ionization}  

To model the ionized and affinitive processes, we shall consider how to remove and/or add one quasielectron to excite the carrier. For convenience, first I focus on the one-bond system in subsection IV-A. The assumptions about the one-bond Hamiltonian $H _{b}$ are discussed in Appendix A. Secondly, I consider only the change of the 4-level subsystem $\eta ^{\prime}$ in subsection IV-B to add and/or remove one quasielectron in the compound system, and discuss the excited carrier near the covalent limit to see the importance of bonding coefficients.  

\section*{IV-A Electron affinity and ionization of the one-bond system}

The one-bond system in Fig. 1 (a) is taken as 4-level dimer to model the ionic-covalent bonding. When one quasielectron is removed from such a dimer, the remained one occupying $| 1 \rangle \otimes | \sigma \rangle $ in Fig. 2 (b) has no correlated partner. Because atom A has the higher electro-negativity, we can approximate the remained quasielectron by $\rho ^{(1)}$ in Eq. (11) and obtain the energy $E_{b} ^{(-)} = tr H _{sb} \rho ^{(1)}$ as $| \Psi _{b} \rangle$ becomes the uncorrelated one-electron state $| \Psi _{b,1, \uparrow } \rangle = c _{ A  \uparrow } ^{\dag} | 0 \rangle$ or $| \Psi _{b,1, \downarrow } \rangle = c _{ A  \downarrow } ^{\dag} | 0 \rangle$. On the other hand, there are three quasielectrons when we change $| \Psi _{b} \rangle$ to a three-electron state, which corresponds to the left-hand side of Fig. 2(c), by the affinitve process. We can see from Fig. 2 (c) that such a three-electron state is equivalent to the one-hole state because there are only 4 levels. The remained quasihole is located at $| \bar{1} \rangle \otimes | \sigma \rangle$, and its spatial part $| \bar{1} \rangle$ can be approximated by $ | B \rangle $ in $\mathbb{C}^{2} _{\Omega}$ because atom $B$ is of the lower electro-negativity. So we can take $| \Psi _{b,3, \sigma } \rangle = c _{ B \sigma} | Fb \rangle \propto c _{ B \underline{ \sigma } } ^{ \dag } c _{ A \sigma } ^{\dag} c _{ A \underline{\sigma} } ^{\dag} | 0 \rangle  \propto c _{ L \underline{\sigma} } ^{\dag} c _{ A \sigma } ^{\dag} c _{ \bar{L} \underline{\sigma} } ^{\dag} | 0 \rangle$ as the wavefunction of the three-electron or one-hole state for $\sigma = \uparrow$ or $\downarrow$, and the added electron is located at the spatial ket $| \bar{L} \rangle \in \mathbb{C}^{2} _{\Omega}$. Here $ \underline{ \sigma } = - \sigma$, $c _{ \bar{L} \underline{ \sigma } } $ is to annihilate the electron at $ | \bar{L} \rangle \otimes | -  \sigma \rangle $, and $| Fb \rangle$ denotes the four-electron state for the filled one-bond system. The state $| \Psi _{b,3, \sigma } \rangle$ is uncorrelated, and its density matrices for $\sigma$ and $- \sigma $ are $\rho ^{(1)} = | A \rangle \langle A |$ and $ \rho ^{(2)} _{+} = | A \rangle \langle A | +  | B \rangle \langle B | = | L \rangle \langle L | +  | \bar{L} \rangle \langle \bar{L} |$, respectively. Hence the energy for the one-bond system becomes $E ^{(+)} _{b} =tr (\rho ^{(1)} +  \rho ^{(2)} _{+})H_{sb} + \int _{ {\bf r} ^{ \prime },{\bf r} ^{\prime \prime} \in \Omega} d ^{3} r ^{\prime} d ^{3} r ^{\prime \prime} U ( | {\bf r} ^{\prime} - {\bf r} ^{\prime \prime} | ) \langle {\bf r} ^{\prime} | \rho ^{(1)} | {\bf r} ^{\prime} \rangle \langle {\bf r} ^{\prime \prime} | \rho ^{(2)} _{+} | {\bf r} ^{\prime \prime} \rangle$ after we add one electron.   

By taking $ \rho ^{(2)} _{-} = d ^{(1)} _{ \pm } = d ^{(2)} _{ \pm } =0$, we can rewrite $E^{( \pm )} _{b}$ as 
\begin{eqnarray}
E_{b} ^{( \pm )}=  tr (\rho ^{(1)} + \rho ^{(2)} _{ \pm }) H_{sb} + \int _{ {\bf r} ^{\prime},{\bf r} ^{\prime \prime} \in \Omega} d ^{3} r ^{\prime} d ^{3} r ^{\prime \prime} U ( | {\bf r} ^{\prime} - {\bf r} ^{\prime \prime} | ) \langle {\bf r} ^{\prime} | \rho ^{(1)} | {\bf r} ^{\prime} \rangle \langle {\bf r} ^{\prime \prime} | \rho ^{(2)} _{ \pm } | {\bf r} ^{\prime \prime} \rangle +
\end{eqnarray}
\[
 tr H_{sb} d ^{(1)} _{ \pm } + \int _{ {\bf r} ^{\prime},{\bf r} ^{\prime \prime} \in \Omega} d ^{3} r ^{\prime} d ^{3} r ^{\prime \prime} U ( | {\bf r} ^{\prime} - {\bf r} ^{\prime \prime} | ) \langle {\bf r} ^{\prime} | \rho ^{(1)} | {\bf r} ^{\prime} \rangle \langle {\bf r} ^{\prime \prime} | d ^{(1)} _{ \pm } | {\bf r} ^{\prime \prime} \rangle + \text{ \ \ \ \ \ }
\]
\[
\int _{ {\bf r} ^{\prime},{\bf r} ^{\prime \prime} \in \Omega} d ^{3} r ^{\prime} d ^{3} r ^{\prime \prime} U ( | {\bf r} ^{\prime} - {\bf r} ^{\prime \prime} | ) \langle {\bf r} ^{\prime} | \rho ^{(1)} | {\bf r} ^{\prime \prime} \rangle \langle {\bf r} ^{\prime \prime} | d ^{(2)} _{ \pm } | {\bf r} ^{\prime} \rangle. \text{ \ \ \ \ \ \ \ \ \ \ \ \ \ \ \ \ \ \ \ \ \ \ \ \ \ }
\]
The above equation can be obtained from Eq. (13) by substituting $ \rho ^{(2)} _{ \pm } $, $ d ^{(1)} _{ \pm }$, and $ d ^{(2)} _{ \pm } $ for $ \rho ^{(2)}$, $ d ^{(1)} $, and $ d ^{(2)} $, respectively.  The meaning of $d ^{(1)} _{ \pm } = d ^{(2)} _{ \pm } =0$ is that there is no fluctuating charge or covalent correlation after the ionized and affinitive processes. The remained quasielectron in Fig. 2 (b) has no correlated partner, so it is natural that  $d ^{(1)} _{ - } = d ^{(2)} _{ - } =0$. On the other hand, only one quasihole is left at the right-hand side of Fig. 2 (c), and the corresponding one-hole state should be similar to the one-electron state in Fig. 2 (b) based on the electron-hole symmetry. Hence it is reasonable that $d ^{(1)} _{ + } = d ^{(2)} _{ + } =0$. 

\section*{IV-B Electron affinity and ionization of the compound system}

In subsection III-B, the quasielectron system to model the considered compound is decomposed into 4-level subsystems, as shown in Fig. 3 (a). To remove (add) one quasielectron from (to) the subsystem characterized by $| \eta ^{ \prime } \rangle \in \{ | \eta _{j} \rangle \}$, I note that $| \eta ^{\prime } \rangle \otimes | A \rangle$ and $| \eta ^{\prime } \rangle \otimes | B \rangle$ in subsystem $\eta ^{\prime}$ serve as $| A \rangle$ and $|B \rangle$ in the one-bond system, respectively. Therefore, we shall introduce $\rho ^{(2)} _{ \eta ^{ \prime} , - } =0$ and $\rho ^{(2)} _{ \eta ^{ \prime} , + } = | \eta ^{ \prime } \rangle \langle \eta ^{ \prime } | \otimes ( | A \rangle \langle A | + | B \rangle \langle B |) $ for the electron ionization and affinity just as how we introduce $\rho ^{(2)} _{  - }$ and $\rho ^{(2)} _{+} $ according to the electronegativities in subsection IV-A. The ionized quasielectron is removed from $| \eta ^{\prime } \rangle \otimes | L \rangle$ while one quasielectron enters $| \eta ^{\prime } \rangle \otimes | \bar{L} \rangle$ in the affinitive process, and the removed/added charge in the j-th chemical bond equals $| \langle w _{j} | \eta ^{\prime} \rangle | ^{2}$. Together with $\rho ^{(1)} _{ \eta ^{\prime}} $, the matrix $\rho ^{(2)} _{ \eta ^{ \prime} , - } $ corresponds to the one-electron state in Fig. 2 (b) while $\rho ^{(2)} _{ \eta ^{ \prime} , + } $ corresponds to the three-electron or one-hole state in Fig. 2 (c). Because the one- and three-electron states are both uncorrelated, we shall take $d ^{(1)}  _{ \eta ^{\prime} , \pm } = d ^{(2)}  _{ \eta ^{\prime} , \pm } =0$ to remove the correlation contribution of the subsystem $ \eta ^{\prime}$. If each subsystem characterized by $ \eta _{j} \neq \eta ^{\prime} $ remains unchanged, as shown in Fig. 3 (b), we shall replace $\rho ^{( \text{II} )}$, $d ^{( \text{I} )}$, and $d ^{( \text{II} )}$ by $ \rho ^{( \text{II} )} _{ \eta ^{\prime} , \pm } = \sum _{ \eta _{j} \neq \eta ^{\prime} } \rho ^{(2)} _{ \eta _{ j } } +\rho ^{(2)} _{ \eta ^{ \prime } , \pm } $,  $d ^{( \text{I} )} _{ \eta ^{\prime} , \pm } = \sum _{ \eta _{j} \neq \eta ^{\prime} } d ^{(1)} _{ \eta _{ j } } +d ^{(1)} _{ \eta ^{ \prime} , \pm }  $, and $d ^{( \text{II} )} _{ \eta _{\prime} , \pm } = \sum _{ \eta _{j} \neq \eta ^{\prime} } d ^{(2)} _{ \eta _{j} }  +d ^{(2)} _{ \eta ^{ \prime} , \pm }  $, respectively. The energy $E _{Cr} $ becomes 
\[
E_{Cr} ^{ ( \eta ^{\prime} , \pm ) } = tr ( \rho ^{ ( \text{I} ) } + \rho ^{ ( \text{II} ) } _{ \eta ^{\prime}, \pm } ) H _{sC} + \sum _{j=1} ^{N} \int _{ {\bf r} ^{\prime} _{ j },{\bf r} ^{\prime \prime} _{ j } \in \Omega _{j} } d ^{3} r ^{\prime} _{ j } d ^{3} r ^{\prime \prime} _{ j } U ( | {\bf r} ^{\prime} _{ j } -  {\bf r} ^{\prime \prime} _{ j } | ) \langle {\bf r} ^{\prime} _{ j } | \rho ^{( \text{I} )} | {\bf r} ^{\prime} _{ j } \rangle \langle {\bf r} ^{\prime \prime} _{ j }  | \rho ^{( \text{II} )} _{ \eta ^{\prime}, \pm} | {\bf r} ^{\prime \prime} _{ j }  \rangle  +
\]
\begin{eqnarray}
trH _{sC} d^{ ( \text{I} )} _{ \eta ^{\prime}, \pm} + \sum _{j=1} ^{N} \int _{ {\bf r} ^{\prime} _{ j },{\bf r} ^{\prime \prime} _{ j } \in \Omega _{j} } d ^{3} r ^{\prime} _{ j } d ^{3} r ^{\prime \prime} _{ j } U ( | {\bf r} ^{\prime} _{ j } - {\bf r} ^{\prime \prime} _{ j } | ) \langle {\bf r} ^{\prime} _{ j } | \rho ^{( \text{I} )} | {\bf r} ^{\prime} _{ j } \rangle  \langle {\bf r} ^{\prime \prime} _{ j }  | d ^{( \text{I} )} _{ \eta ^{\prime}, \pm} |{\bf r} ^{\prime \prime} _{ j }  \rangle  +
\end{eqnarray}
\[
 \sum _{j=1} ^{N} \int _{ {\bf r} ^{\prime} _{ j },{\bf r} ^{\prime \prime} _{ j } \in \Omega _{j} } d ^{3} r ^{\prime} _{ j } d ^{3} r ^{\prime \prime} _{ j } U ( | {\bf r} ^{\prime} _{ j } -  {\bf r} ^{\prime \prime} _{ j } | ) \langle {\bf r} ^{\prime} _{ j } | \rho ^{( \text{I} )} | {\bf r} ^{\prime \prime } _{ j } \rangle \langle {\bf r} ^{\prime \prime} _{ j }  | d ^{( \text{II} )} _{ \eta ^{\prime}, \pm} | {\bf r} ^{\prime} _{ j }  \rangle \text{ \ \ \ \ \ \ \ \ \ \ \ \ \ \ \ \ \ \ } 
\]
after we change the number of electrons in subsystem $\eta ^{\prime} $. 

An effective carrier is excited in the ionized/affinitive process, and we can obtain its excitation energy by calculating the difference $E_{Cr} ^{ ( \eta ^{\prime} , \pm ) } -E_{Cr}$ based on Eqs. (18) and (20). It is convenient to rewrite $\rho ^{( \text{II} )} _{ \eta ^{\prime} , \pm }$, $d ^{( \text{I} )} _{ \eta ^{\prime} ,\pm }$, and $d ^{( \text{II} )} _{ \eta ^{\prime} ,\pm }$ as
\begin{eqnarray}
\begin{cases}
\rho ^{( \text{II} )} _{ \eta ^{\prime} , + }  = \rho ^{( \text{II} )} + | \eta ^{ \prime } \rangle \langle \eta ^{ \prime } | \otimes | \bar{L} \rangle \langle \bar{L} | \\
\rho ^{( \text{II} )} _{ \eta ^{\prime} , - } = \rho ^{( \text{II} )} -\rho ^{(2)} _{ \eta ^{\prime} } \\
d ^{( \text{I} )} _{ \eta ^{\prime} ,\pm }  = ( \sqrt{2} -1) [ \rho ^{( \text{I} )} \rho ^{( \text{II} )} _{ \eta ^{\prime} , -}  ( I - \rho ^{( \text{I} )}) +   ( I - \rho ^{( \text{I} )}) \rho ^{( \text{II} )}  _{ \eta ^{\prime} , - } \rho ^{( \text{I} )} ]   \\  
d ^{( \text{II} )} _{ \eta ^{\prime} ,\pm } =  ( I - \rho ^{( \text{I} )})  \rho ^{( \text{II} )}  _{ \eta ^{\prime} , - } ( I - \rho ^{( \text{I} )}) \text{ \ \ \ \ \ \ \ \ \ \ \ \ \ \ \ \ \ \ \ \ \ \ \ \ \ \ \ \ }
\end{cases},
\end{eqnarray}
in Eq. (20) to obtain the result irrelevant to $| \eta _{j} \rangle $ for all $ \eta _{j} \neq \eta ^{\prime}$. Direct calculation yields
\begin{eqnarray}
E_{Cr} ^{ ( \eta ^{\prime} , \pm ) } -E_{Cr}=\langle \eta ^{ \prime} | H _{\pm } | \eta ^{ \prime } \rangle + E _{ b} ^{ (\pm)} - E _{b} \text{ with }  
\begin{cases}
H_{+} = tr ^{ \prime } H _{hop} (| \bar{L} \rangle \langle \bar{L} | - d ^{(1)}) \\
H_{-} = - tr ^{ \prime } H _{hop} (| L \rangle \langle L | + d ^{(1)}) 
\end{cases},
\end{eqnarray}
where $tr ^{\prime} $ denotes the trace with respect to $\mathbb{C} ^{2} _{ \Omega}$. Near the ionic limit, it is shown in Appendix C that $E_{Cr} ^{ ( \eta ^{\prime} , \pm ) } -E_{Cr}$ can be close to the energy difference obtained by considering the coupled-cluster corrections after we improve my model based on Eq. (25).  

The operators $| \bar{L} \rangle \langle \bar{L} |$,  $| L \rangle \langle L |$, and $ d ^{(1)}$ in Eq. (22) include the bonding coefficients $\alpha _{i}$ and $\alpha _{c}$, which depend on the repulsive strength of the e-e interaction potential $\hat{U}$ as mentioned in Appendix B. Therefore, we can obtain the interaction-dependent electron ionization and affinity for the excited carrier. To obtain the quantitative results, I consider the nearest-neighbor hopping in the zincblende-structure crystal, in which the atom $A$ located at ${\bf R} _{A} = n _{1} (l_{c}/2, l_{c}/2, 0) + n _{2} (0, l_{c}/2,l_{c}/2) + n _{3} (l_{c}/2, 0, l_{c}/2)$ is accompanied by the atom $B$ at  ${\bf R} _{B} = {\bf R} _{A} + (l_{c}/4, l_{c}/4,  l_{c}/4) $. Here $n _{1}$, $n _{2}$, and $n _{3}$ are integers, and $l _{c}$ is the length of the crystal lattice. For convenience, I denote ${\bf n} = (n _{1}, n _{2}, n _{3}) $ for the parameters $n _{1}$, $n _{2}$, and $n _{3}$ of ${\bf R} _{A}$, and take $ m =1 \sim 4$ to parameterize the 4 ionic-covalent bonds around the same atom $A$ such that each $| w _{j} \rangle$ can be re-parameterized as $| w _{ {\bf n}, m } \rangle$. Assume that the hopping coefficients equal $t _{A}$ and $t _{B}$ for the adjacent A- and B-site orbitals, respectively. By adding one quasielectron to the s-like Bloch-type orbital $ | \eta ^{\prime} \rangle =\frac{1}{2 \sqrt{N} }  \sum _{ {\bf n} , m } e ^{ i {\bf k} \cdot {\bf R}_{A} } | w_{ {\bf n}, m } \rangle \in \mathbb{C} ^{N}$ near the covalent limit, we can obtain 
\begin{eqnarray}
E_{Cr} ^{ ( \eta ^{\prime} , + ) } -E_{Cr}=3  t _{A} | \alpha _{c} |  ^{2} + E _{b} ^{(+)} - E _{b} -\gamma  \varepsilon _{dis} ( {\bf k}) \text{ with } 
\end{eqnarray}
\[
 \varepsilon _{dis} ( {\bf k}) = cos \frac{ l _{c} k _{x} }{2}  cos \frac{ l _{c} k _{y} }{2} +  cos \frac{ l _{c} k _{y} }{2}  cos \frac{ l _{c} k _{z} }{2}  +  cos \frac{ l _{c} k _{z} }{2}  cos \frac{ l _{c} k _{x} }{2} 
\]
as the energy dispersion curve \cite{Grosso} for the excited carrier in the tight-binding scheme. Here $ \gamma = - | \alpha _{i}| ^{2} t _{B} $ serves as the overlap energy \cite{Kittel}, ${\bf k}= ( k _{x} , k _{y}, k _{z}) $ denotes the wavevector, and  $\varepsilon _{dis} ( {\bf k})$ can be obtained by considering the twelve nearest-neighbor vectors \cite{Grosso}. The effective mass $m ^{*} ({\bf k})$ \cite{Grundmann} follows $ [ m ^{*-1}({\bf k}) ] _{i,j} =  - \frac{ \gamma }{\hbar ^2} \frac{ \partial ^{2} }{\partial k_{i} \partial k _{j}} \varepsilon _{dis} ( {\bf k}) \propto |\alpha _{i}| ^{2} = 1- |\alpha _{c}| ^{2} $ at each ${\bf k}$, which reveals the importance of the bonding coefficients to the excited carrier. In addition, the carrier becomes immobile in the covalent limit because the bandwidth equals zero as $| \alpha _{i} | = 0$. The coefficients $\alpha _{i}$  and $\alpha _{c}$ are determined by the e-e repulsive strength as mentioned in Appendix B, so $m ^{*} ({\bf k})$ depends on the e-e interaction potential $\hat{U}$ in my model. The zero bandwidth in the covalent limit is due to the lack of the double occupancy at half filling under the strong repulsive strength of $\hat{U}$, which induces the Mott insulating behaviors in some AF systems \cite{Lee1,Liu}. 

\section{Discussion} 

In the last three sections, the non-negative function $U(| {\bf r} |)$ is taken into account to introduce $\hat{U}$ without considering the inter-bond e-e correction. Actually there should exist the inter-bond e-e energy $\frac{1}{2} \sum _{  j_{1} \neq j_{2} } \int _{ {\bf r} _{1} \in \Omega _{ j _{1} } } d ^{3} r _{1}  \int _{ {\bf r} _{2} \in \Omega _{ j _{2} } } d ^{3} r _{2} U ^{\prime} (| {\bf r} _{1} - {\bf r} _{2} |) Q ({\bf r} _{1})Q({\bf r} _{2})$ in the quasielectron system to model the considered compound, where $U ^{\prime}$ represents the long-range e-e correction. Because the electron density $Q({\bf r} _{j} )= 2 \langle {\bf r} _{j} | \rho _{sC} | {\bf r} _{j} \rangle$ at ${\bf r} _{j} \in \Omega _{j}$ in the $j$-th bond and $\rho _{sC} = \frac{1}{2}( \rho ^{ ( \text{I} ) } + \rho ^{ ( \text{II} ) } + d ^{ ( \text{I} ) }) $, we shall include
\[
\text{(i) } \frac{1}{2} \sum _{ q = \text{I}, \text{II} } \text{ } \sum_{ q ^{\prime} = \text{I}, \text{II} } \text{ } \sum _{  j_{1} \neq j_{2} } \int _{ {\bf r} _{1} \in \Omega _{ j _{1} } } d ^{3} r _{1}  \int _{ {\bf r} _{2} \in \Omega _{ j _{2} } } d ^{3} r _{2} U ^{\prime} (| {\bf r} _{1} - {\bf r} _{2} |) \langle {\bf r} _{1} | \rho ^{ (q) } | {\bf r} _{1} \rangle \langle {\bf r} _{2} | \rho ^{(q^{\prime})} | {\bf r} _{2} \rangle 
\]
\begin{eqnarray}
\text{(ii) \  }  \sum _{  q = \text{I}, \text{II} } \text{ } \sum _{  j_{1} \neq j_{2} } \int _{ {\bf r} _{1} \in \Omega _{ j _{1} } } d ^{3} r _{1}  \int _{ {\bf r} _{2} \in \Omega _{ j _{2} } } d ^{3} r _{2} U ^{\prime} (| {\bf r} _{1} - {\bf r} _{2} |) \langle {\bf r} _{1} | \rho ^{ (q) } | {\bf r} _{1} \rangle \langle {\bf r} _{2} | d ^{ (I) } | {\bf r} _{2} \rangle \text{ \ \ \ \ \ \ } 
\end{eqnarray}
\[ 
\text{(iii) } \frac{1}{2} \sum _{  j_{1} \neq j_{2} } \int _{ {\bf r} _{1} \in \Omega _{ j _{1} } } d ^{3} r _{1}  \int _{ {\bf r} _{2} \in \Omega _{ j _{2} } } d ^{3} r _{2} U ^{\prime} (| {\bf r} _{1} - {\bf r} _{2} |) \langle {\bf r} _{1} | d^{( \text{I} )} | {\bf r} _{1} \rangle \langle {\bf r} _{2} | d^{( \text{I} )} | {\bf r} _{2} \rangle \text{ \ \ \ \ \ \ \ \ \ \ \ \ \ \ \ \ } 
\]
for the inter-bond e-e correction. Since each chemical bond in the compound system is identical to the one-bond system discussed in subsection III-A, the charge fluctuation $\langle {\bf r} _{j} | d ^{( \text{I})} | {\bf r} _{j} \rangle$ at ${\bf r} _{j} \in \Omega _{j}$ for any $j$ has no contribution to the total electron charge just as $\langle {\bf r} | d ^{(1)} | {\bf r} \rangle$. I note that the deviation $\delta \rho$ responsible for the density-density interaction \cite{You} in the quantum Hall theory \cite{Zhang,DHLee} also has no contribution to the total charge, and  Eq. (24)-(iii) shows the universality of such interaction. The charge fluctuation due to $ d ^{( \text{I} )}$ may interact with the charge density given by $\rho ^{( \text{I} )}$ and $\rho ^{( \text{II} )}$, and Eq. (24)-(ii) just provides the corresponding energy together with Eq. (16)-(ii). Equations (16)-(i) and (24)-(i) yield the Hartree-potential energy resulting from $\rho ^{( \text{I} )}$ and $\rho ^{( \text{II} )}$, and we shall include the Fock-potential term \cite{Nelson} because of the lack of the self-interaction in Eq. (16)-(i).  

When the compound is an ideal crystal following the periodic boundary condition, it is important to consider the case that each $ |  \eta _{j} \rangle \in \mathbb{C} ^{N}$ in Fig. 3 (a) corresponds to one Bloch-type function. The Bloch wavefunction, which consists of its Bloch-type part in $ \mathbb{C} ^{N}$ and the bonding part in $ \mathbb{C} ^{2} _{ \Omega}$, can be extended to the form $e ^{ i {\bf k} \cdot {\bf r} } u _{ {\bf k} } ({\bf r}) $ \cite{Kittel} to include the small density due to the electron tails in the shaded region in Fig. 1 (c) as $H _{ sC} $ is replaced by $\frac{{\bf p}^{2}}{2m_{0}} +V _{cr}({\bf r})$. Here $V_{cr}$ denotes the one-electron periodic potential in the crystal, $m _{0} $ is the electron mass in vacuum, ${\bf p}$ is for the momentum operators, and $u _{ {\bf k} } ( {\bf r})$ represents the periodic part of the corresponding Bloch state. To determine the hopping coefficients $t _{ j \xi, j^{\prime} \xi ^{\prime}}$ in Eq. (17), principally we can transform the Bloch wavefunctions to the Wannier ones \cite{Marzari}, which serve as the localized atomic orbitals in the tight-binding model \cite{Mahan}. For the well-developed ionic-covalent bonds, in Fig. 1 (c) the electron density in the shaded region must be so low that the bonding electrons almost concentrate in the red region, where the high density induces the bonding correlation representing by Eq. (15). After introducing the inter-bond hopping, therefore, we may approximate the $j$-th bond's Wannier function as zero outside $\Omega _{j} $ for all $j$ to calculate the e-e energy. Actually Eqs. (18) and (20) can be valid in the systems composed of different ionic-covalent dimers such as the those in the chalcopyrite-structure \cite{Grundmann} compound when all the bonds are well-separated, as shown in Appendix D. The Wannier basis, however, depends on the gauge freedom \cite{Marzari} and is not unique. When the ionic-covalent bonds in the compound are not identical, the decomposition in Fig. 3 (a) can become invalid because of the non-constant bonding coefficients. More studies are necessary to clarify how to exactly include the bonding correlation beyond the compound model developed in subsections III-B and IV-B. 

It is shown in subsection IV-B that bandwidth can become zero because of the strong e-e repulsive strength, which is responsible for the Mott insulator in some AF systems \cite{Lee1,Liu}. It is known that the random fields \cite{Kirsc,Erdos,Lee2,Huang2} modeled by a family of parameterized Hamiltonians can result in the disorder leading to different insulators, and both the disorder and e-e interaction effects have been observed in the quantum Hall systems \cite{TYH,Wang}. The transition between insulating phases has been studied by considering the disordered interacting systems. \cite{Byczuk,Braganc} To include a Hamiltonian family, we may replace $H_{sC}$ by the random-matrix set $H_{sC} ^{\omega}$ parameterized by $\omega$ and consider $\sum _{\omega} (\rho ^{( \text{I} )}_{\omega} e _{1,\omega} e _{1,\omega} ^{\dag} + \rho ^{( \text{II} )}_{\omega} e _{2,\omega} e _{2,\omega} ^{\dag})$. Here the set $\{ e _{1, \omega} , e _{2,\omega} \}$ is an orthonormal one in the corresponding vector space, and for each $\omega$ the matrices $\rho ^{( \text{I} )}_{\omega}$ and $\rho ^{( \text{II} )}_{\omega}$ serve as $\rho ^{( \text{I} )}$ and $\rho ^{( \text{II} )}$. If the one-bond system in Fig. 1 (a) is asymmetric with respect to the bonding axis, the rotation centered on such an axis is important to the mapping ${\bf R} _{j}$ in subsection III-B for each $j$ and we need to introduce the parameter $\omega$ for the rotation degrees of freedom \cite{Huang3}. In the Born-Oppenhemier method \cite{Bohm} (BOM), we also need to consider a family of Hamiltonians to determine the electron wavefunctions parameterized by the relative position of the nuclei. For any two $2^{n} \times 2^{n}$ matrices $\rho ^{\prime} _{ea}$ and $\rho ^{\prime \prime} _{ea}$, actually we can construct a $ 2 ^{n+1} \times 2 ^{n+1}$ matrix $\rho ^{\prime \prime \prime} _{ea} = \rho ^{\prime} _{ea} \otimes e _{1} e _{1} ^{\dag} + \rho ^{\prime \prime} _{ea} \otimes e _{2} e _{2} ^{\dag}$  and take Eq. (8) as the case for $n=0$, where the integer $n$ is non-negative. By this way we can construct AF-type quasielectrons with $2^{n+1}$ components for the Hamiltonian family parameterized by $\omega = 1 \sim 2 ^{n}$. The multiple-component orbitals can be used to include the multiple CCDs, which are briefly discussed in Appendix B after including $| \Psi _{BB} \rangle$, in the quasielectron space \cite{Huang3}. I note that the multiple-component functions are introduced to develope the vector bundles \cite{Bohm,Friedman,Banks}. While the hole components \cite{Huang1} do not appear in the density matrices in my ionic-covalent model, they may become important when both the particle-particle and particle-hole channels \cite{Yu} are taken into account for the Bogoliubov-BCS quasiparticles. By considering the fractal structures \cite{Schwalm} to extend such quasiparticles \cite{Huang1,Huang4}, in fact, we can obtain the form of Eq. (2) from the electron components of the extended ones.    

\section{Summary}

The extended AF quasielectrons are introduced for the compound where the ionic-covalent bonds are identical to the one-bond dimer. The density-fluctuation and covalent-correlation operators are constructed for the bonding correlation, and my quasielectron model shows the universality of the density-density interaction. The quasielectron system is decomposed into the 4-level subsystems for the electron ionization and affinity in the compound, and such a model can be supported by the coupled-cluster theory near the ionic limit. For the ideal crystal, each subsystem may correspond to one Bloch-type function. By considering the nearest-neighbor hopping in the zincblende-structure crystal, we can see the importance of the bonding coefficients to the effective mass near the covalent limit.  

\section*{Acknowledgment}

The author thanks Profs. I.-H. Tsai, Keh-Ning Huang, and Hrong-Tzer Yau for the valuable discussions about the vector bundles, coupled-cluster corrections, and random-matrix theory, respectively.

\section*{Appendix A}

For the one-bond system in Fig. 1 (a), I assume that the distance $\ell$ separating the peaks of $| \langle {\bf r } | A \rangle | ^{2}$ and $| \langle {\bf r } | B \rangle | ^{2}$ in Fig. 4 is longer than the interacting length $a$ of $\hat{U}$. In addition, assume that $\int _{ {\bf r} ^{\prime} \in \Omega} d ^{3} r ^{\prime} \int _{ {\bf r ^{\prime \prime} } \in \Omega} d^{3} r ^{ \prime \prime}  U (|{\bf r} ^{\prime} -{\bf r} ^{ \prime \prime } |) | \langle {\bf r } ^{\prime} | A \rangle | ^{2} | \langle {\bf r ^{\prime \prime } } | A \rangle | ^{2} \sim \int _{ {\bf r} ^{\prime}  \in \Omega} d ^{3} r ^{\prime}  \int _{ {\bf r ^{\prime \prime } } \in \Omega} d^{3} r ^{ \prime \prime} U (|{\bf r} ^{\prime} -{\bf r} ^{ \prime \prime} |) | \langle {\bf r }^{\prime} | B \rangle | ^{2} | \langle {\bf r ^{\prime \prime} } | B \rangle | ^{2} \sim U _{0} >0$, where the positive parameter $U _{  0} $ represents the repulsive strength of $\hat{U}$. Hence the effective e-e potential $\hat{U}$ in Eqs. (9) and (10) is dominated by $U _{0} c _{A \uparrow} ^{\dag} c _{A \uparrow} c _{A \downarrow} ^{\dag} c _{A \downarrow} + U _{0} c _{B \uparrow} ^{\dag} c _{B \uparrow} c _{B \downarrow} ^{\dag} c _{B \downarrow}  \equiv \hat{U} _{0} $, which corresponds to the intrasite Coulomb repulsion \cite{Matlak1,Matlak2} in the t-U model. Moreover, I assume that the difference $\varepsilon _{B} -  \varepsilon_{A} $ is high enough for us to take $\sum _{\sigma} (t _{AB} c _{A \sigma} ^{\dag} c _{B \sigma} + t _{AB} ^{*} c _{B \sigma} ^{\dag} c _{A \sigma} ) + \hat{U}-  \hat{U}_{0}$ as the perturbation part of $H _{b}$.       

Under the above assumptions, the remained quasielectron and quasihole in the one-bond system are roughly located at $| A \rangle$ and $| B \rangle$ as the one-bond wavefunction becomes the one- and three-electron ground states, respectively. Hence the spatial ket $| 1 \rangle $ of the occupied level in Fig. 2 (b) is close to $| A \rangle $ when the 4-level dimer in Fig. 2 represents such a one-bond system, and the ket $| \bar{1} \rangle$ for the quasihole at the right-hand side of Fig. 2 (c) can be approximated by $| B \rangle$. In addition, we may neglect the small change on $\rho ^{(1)}$ as one electron is removed/added. For the one- and three-electron states of subsystem $ \eta ^{\prime} $ in Fig. 3 (b), on the other hand, we shall take $| 1 \rangle \sim | \eta ^{\prime} \rangle \otimes | A \rangle $ and $| \bar{1} \rangle \sim | \eta ^{\prime} \rangle \otimes | B \rangle$  in Figs. 2 (b) and (c), respectively, and approximate $\rho ^{(1)} _{\eta ^{\prime} }$ as $( | \eta ^{\prime} \rangle \otimes | A \rangle)( \langle \eta ^{\prime} | \otimes \langle A |)$. We can tune $\rho ^{(1)}$ and $\rho ^{(1)} _{ \eta ^{\prime} }$ in section IV to modify the orbitals of the remained quasiparticles in the corresponding spaces $\mathbb{C} ^{2} _{\Omega}$ and $| \eta ^{\prime} \rangle \langle \eta ^{\prime} | \otimes \mathbb{C} ^{2} _{\Omega}$. The charge background due to the half-filled subsystems, which are characterized by $\eta _{j} \neq \eta ^{\prime}$ in Fig. 3 (b), should be taken into account to perform the modification for the compound system.  

\section*{Appendix B}

In the one-bond system in Fig. 1 (a), the wavefunction $| \Psi _{b} ^{SCF} \rangle = ( \sqrt{1- | \lambda _{1} | ^{2} } c _{ A \uparrow} ^{\dag} + \lambda _{1}  c _{ B \uparrow} ^{\dag} )  ( \sqrt{1- | \lambda _{1} | ^{2} } c _{ A \downarrow} ^{\dag} + \lambda _{1}  c _{ B \downarrow} ^{\dag} ) | 0 \rangle$ can serve as the effective SCF state at half filling near the ionic limit if the small parameter $\lambda _{1}$ is determined by minimizing $\langle  \Psi _{b} ^{SCF} |  H _{b} | \Psi _{b} ^{SCF} \rangle$. The wavefunction $| \Psi _{b} ^{SCF} \rangle$ is a linear combination of $| \Psi _{i} \rangle$, $| \Psi _{c} \rangle$, and $| \Psi _{BB} \rangle$, so principally we should take $| \Psi _{BB} \rangle$ into account in addition to the ionic and covalent parts. The ket $| \Psi _{b} ^{CCD} \rangle = ( \sqrt{1- | \lambda _{1} | ^{2} } c _{ B \uparrow} ^{\dag} - \lambda _{1} ^{*} c _{ A \uparrow} ^{\dag} )  ( \sqrt{1- | \lambda _{1} | ^{2} } c _{ B \downarrow} ^{\dag} - \lambda _{1} ^{*}  c _{ A \downarrow} ^{\dag} ) | 0 \rangle$ is the only allowed CCD for the bonding electrons. When Brillouin theorem \cite{Manninen,Pople} is valid near the ionic limit, the coupled-cluster method is applicable and we may take $| \Psi _{b} ^{Br} \rangle =  \sqrt{1- | \lambda _{2} | ^{2} } | \Psi _{b} ^{SCF} \rangle + \lambda _{2} | \Psi _{b} ^{CCD} \rangle $ as $| \Psi _{b} \rangle $ to model the ground state. Here the small parameter $\lambda _{2}$ is determined by minimizing $\langle \Psi _{b} ^{Br} | H _{b} | \Psi _{b} ^{Br} \rangle$. The single substitution \cite{Pople} is neglected in $| \Psi _{b} ^{Br} \rangle $.

While Brillouin theorem may become invalid, we have the bonding wavefunction $ | \Psi _{b} \rangle = \tau _{i} | \Psi _{i} \rangle + \tau _{c} | \Psi _{c} \rangle+ \tau _{BB} | \Psi _{BB} \rangle $ \cite{Prasad,Havenith} in general when the one-bond system is half-filled. Here $ \tau _{i} $, $ \tau _{c} $, and $ \tau _{BB} $ are the coefficients satisfying $ | \tau _{i} | ^{2} + | \tau _{c} | ^{2} + | \tau _{BB} | ^{2} =1$. The energies of $| \Psi _{i} \rangle$ and $| \Psi _{c} \rangle$ are close to $2 \varepsilon _{A} + U _{0}$ and $\varepsilon _{A}+\varepsilon _{B}$, respectively, and are both lower than the energy of $| \Psi _{BB} \rangle$ under the assumptions mentioned in Appendix A. Hence $| \tau_{BB} | $ should be small, and we can approximate $\tau _{i}$ and $\tau _{c}$ as the coefficients  $\alpha _{i}$ and $\alpha _{c}$ in Eq. (1) if it is suitable to neglect the small contribution of $ | \Psi _{BB} \rangle$. The bonding wavefunction $| \Psi _{b} \rangle \rightarrow | \Psi _{i} \rangle$ near the ionic limit as the repulsive strength $ U_{0} << \varepsilon _{B}- \varepsilon _{A}$. With increasing the e-e repulsive strength, $| \alpha _{i} |$  decreases and $\alpha _{c}$ becomes significant. The wavefunction $|\Psi _{b} \rangle \rightarrow |\Psi _{c} \rangle$ near the covalent limit when $U _{0} >> \varepsilon _{B}- \varepsilon _{A} $, under which the double occupancy is forbidden. 

When it is inappropriate to neglect $| \Psi _{BB} \rangle$,  we can perform the orbital transformation
\begin{eqnarray}
\begin{cases}
c _{ A^{\prime} \sigma } =\sqrt{ 1 - | \lambda _{3} | ^{2} } c _{ A \sigma } + \lambda _{3} c _{ B \sigma } \\
c _{ B^{\prime} \sigma } = - \lambda _{3} ^{*} c _{ A \sigma } + \sqrt{ 1 - | \lambda _{3} | ^{2} } c _{ B \sigma }
\end{cases},
\end{eqnarray}
to rewrite $| \Psi _{b} \rangle$ as $\alpha_{i} ^{\prime} | \Psi _{i} ^{\prime}  \rangle  + \alpha_{c} ^{\prime} | \Psi _{c} ^{\prime}  \rangle + \tau_{BB} ^{\prime} | \Psi _{BB} ^{\prime}  \rangle  $ with $ | \Psi _{i} ^{\prime}  \rangle = c _{ A^{\prime} \uparrow } ^{\dag}  c _{ A^{\prime} \downarrow } ^{\dag} | 0 \rangle  $, $| \Psi _{c} ^{\prime}  \rangle = \frac{1}{ \sqrt{2} } (c _{ A^{\prime} \uparrow } ^{\dag}  c _{ B^{\prime} \downarrow } ^{\dag}  + c _{ B^{\prime} \uparrow } ^{\dag}  c _{ A^{\prime} \downarrow } ^{\dag}) | 0 \rangle$, and $ | \Psi _{BB} ^{\prime}  \rangle = c _{ B^{\prime} \uparrow } ^{\dag}  c _{ B^{\prime} \downarrow } ^{\dag} | 0 \rangle  $. Here $\lambda _{3}$ is a complex number following $0 \leq | \lambda _{3} | \leq 1 $, and $\alpha_{i} ^{\prime}$, $\alpha_{c} ^{\prime}$, and $\tau_{BB} ^{\prime}$ are the coefficients determined by $\tau _{i}$, $\tau _{c}$, $\tau _{BB}$ and $\lambda _{3}$. The bonding wavefunction $| \Psi _{b} \rangle$ becomes $\alpha_{i} ^{\prime} | \Psi _{i} ^{\prime}  \rangle  + \alpha_{c} ^{\prime} | \Psi _{c} ^{\prime}  \rangle$, the ionic-covalent form given by Eq. (1), if  the parameter $\kappa \equiv \lambda ^{*} _{3} / \sqrt{1 - | \lambda ^{*}  _{3} | ^{2} } $ follows $\tau _{i} \kappa ^{2} - \sqrt{2} \tau _{c} \kappa + \tau _{BB} =0$ such that $\tau_{BB} ^{\prime} =0 $.  There are two solutions to $\kappa$, and we can choose the solution with the smaller absolute value while the other one is important to the spontaneous symmetry breaking \cite{Huang3}. To improve my model by including $| \Psi _{BB} \rangle$, we shall replace $|A\rangle$ and $| B \rangle $ by $| A ^{\prime} \rangle \equiv \sqrt{ 1 - | \lambda _{3} | ^{2} } | A \rangle + \lambda _{3} ^{*} | B \rangle $ and $| B ^{\prime} \rangle \equiv - \lambda _{3} | A \rangle + \sqrt{ 1 - | \lambda _{3} | ^{2} } | B \rangle$ such that $\rho ^{(1)} \rightarrow | A ^{\prime} \rangle \langle A ^{\prime} | $ and $\rho ^{(2)} \rightarrow | L ^{\prime} \rangle \langle L ^{\prime} | $ in Eq. (11), where $ | L ^{\prime} \rangle = \alpha _{i} ^{\prime} | A ^{\prime} \rangle +  \alpha _{c} ^{\prime} | B ^{\prime} \rangle$. In addition, the ket $| \bar{L} \rangle$ in subsection III-A should be replaced by $ | \bar{L} ^{\prime} \rangle = \alpha _{c}  ^{\prime \ast}   | A ^{\prime} \rangle -  \alpha _{i} ^{\prime \ast}   | B ^{\prime} \rangle $. Based on the mapping ${\bf R}_{j}$, for Eq. (14) we have $ \rho _{ w _{j} } ^{(1)} = | w _{j} \rangle \langle w _{j} | \otimes |A ^{\prime} \rangle \langle A ^{\prime} |$ and $ \rho _{ w _{j} } ^{(2)} = | w _{j} \rangle \langle w _{j} | \otimes |L ^{\prime} \rangle \langle L ^{\prime} |$, under which $ \rho _{ \eta _{j} } ^{(1)} = | \eta _{j} \rangle \langle \eta _{j} | \otimes |A ^{\prime} \rangle \langle A ^{\prime} |$ and $ \rho _{ \eta _{j} } ^{(2)} = | \eta _{j} \rangle \langle \eta _{j} | \otimes |L ^{\prime} \rangle \langle L ^{\prime} |$. The matrix $\rho ^{(2)} _{+}$ in Eq. (19) and the operator $\rho ^{(2)} _{ \eta ^{\prime} +}$ for subsystem $\eta ^{\prime}$ in subsection IV-B both remain unchanged because $| A \rangle \langle A| + | B \rangle \langle B | = | A ^{\prime}  \rangle \langle A ^{\prime} | + | B ^{\prime} \rangle \langle B ^{\prime} | $. The one- and three-electron states $| \Psi _{b,1,\sigma} \rangle$ and $| \Psi _{b,3,\sigma} \rangle$ in subsection IV-A should be replaced by $| \Psi ^{\prime} _{b,1,\sigma} \rangle \equiv c ^{\dagger} _{ A^{\prime} \sigma } | 0 \rangle$ and $ | \Psi ^{\prime} _{b,3,\sigma} \rangle \equiv c _{ B^{\prime} \sigma } | Fb \rangle $ if we neglect the small change on $\rho ^{(1)}$ when one electron is removed/added. In Eqs. (19) and (20), we can perform the modification discussed at the end of Appendix A by tuning $\rho ^{(1)}$ and $\rho ^{(1)} _{ \eta ^{\prime} }$.   

When we take $|\Psi ^{Br} _{b} \rangle $ as the bonding wavefunction according to the coupled-cluster method near the ionic limit, the coefficient $\tau _{c}$ can be small and become comparable with $\tau _{BB}$. The bonding wavefunction $| \Psi _{b} \rangle $ is dominated by $| \Psi _{i} \rangle$, but we cannot consider only the ionic part to probe the bonding correlation. Therefore, it is important to improve my model near such a limit by using Eq. (25) to include $| \Psi _{BB} \rangle$ in addition to the covalent part if we hope to exactly probe the bonding correlation.      
   
In the BOM, a family of Hamiltonians are taken into account by considering the variation on the positions of the nuclei. While $| \Psi _{b} ^{CCD} \rangle$ is the only allowed CCD in the 4-level dimer for the one-bond system in Fig. 1 (a), it depends on the positions of the nuclei and thus can generate a CCD family $ \{ | \Psi _{b} ^{CCD} (\omega) \rangle \}$. Here the parameter $\omega$ is to parameterize such a family. Multiple CCDs, in fact, can be incorporated in the quasiparticle space by considering the corresponding family \cite{Huang3}, and we may extend the BOM to develop the quasiparticles including both the electron-correlation and nucleus-vibration effects.    

\section*{Appendix C}

In subsection III-B, I consider the compound system where the identical bonds are parametrized by $j$. For convenience, let $c_{jA \sigma}$ and $c_{jB \sigma}$ as the annihilators to remove electrons with the spin orientation $\sigma$ in $| A, j \rangle$ and $| B, j \rangle$, respectively. The compound Hamiltonian ${\cal H}= \sum _{ j \neq j ^{ \prime}} H_{hop} ^{(j, j^{\prime})} + \sum _{ j } (H_{sb} ^{(j)} + \hat{U _{j} }) $, where $ H_{hop} ^{(j, j^{\prime})} =  \sum _{\sigma} \sum _{ \xi , \xi ^{\prime} \in \{ A,B \} } t _{ j \xi , j ^{\prime} \xi ^{\prime} } c_{j \xi \sigma} ^{\dag} c_{  j ^{\prime} \xi ^{\prime} \sigma} $, $H_{sb} ^{(j)} = \sum _{\sigma} (t _{AB} c _{j A \sigma} ^{\dag} c _{ j B \sigma} + t _{AB} ^{*} c _{jB \sigma} ^{\dag} c _{jA \sigma} + \varepsilon _{A}  c _{jA \sigma} ^{\dag} c _{jA \sigma} + \varepsilon _{B}  c _{jB \sigma} ^{\dag} c _{jB \sigma})$, and $\hat{U _{j} } = \int _{ {\bf r} _{1},{\bf r} _{2} \in \Omega _{j} } d ^{3} r _{1} d ^{3} r _{2} U ( | {\bf r} _{1} - {\bf r} _{2} | ) \psi _{ \uparrow } ^{\dag} ({\bf r} _{1}) \psi _{ \uparrow } ({\bf r} _{1}) \psi _{ \downarrow } ^{\dag} ({\bf r} _{2}) \psi _{ \downarrow } ({\bf r} _{2})$. 

Assume that all the hopping coefficients $ t _{ j \xi , j ^{\prime} \xi ^{\prime} }$ are so small that every bond in the considered compound is almost independent and is mapped from the one-bond system in Fig. 1 (a). When $| \Psi ^{Br} _{b} \rangle$ is taken as the one-bond wavefunction near the ionic limit as mentioned in Appendix B, we shall take $ | \Psi ^{Br} _{j} \rangle=  \sqrt{1- | \lambda _{2} | ^{2} } | \Psi _{j} ^{SCF} \rangle + \lambda _{2} | \Psi _{j} ^{CCD} \rangle$ with $| \Psi ^{SCF} _{j} \rangle=c _{j A^{\prime \prime} \uparrow } ^ {\dag}  c _{j A^{\prime \prime} \downarrow } ^ {\dag} | 0 \rangle$ and $| \Psi ^{CCD} _{j} \rangle =  c _{j B^{\prime \prime} \uparrow } ^ {\dag}  c _{j B^{\prime \prime} \downarrow } ^ {\dag} | 0 \rangle$ for the $j$-th bond based on the mapping ${\bf R}_{j}$. Here $c _{j A^{\prime \prime} \sigma } ^ {\dag}= \sqrt{1- | \lambda _{1} | ^{2} } c _{ j A \sigma} ^{\dag} + \lambda _{1}  c _{ j B \sigma} ^{\dagger} $ and $c _{j B^{\prime \prime} \sigma } ^ {\dag} = \sqrt{1- | \lambda _{1} | ^{2} } c _{ j B \sigma} ^{\dag} - \lambda _{1} ^{*} c _{ j A \sigma} ^{\dag} $. We may use Eq. (25) to rewrite $| \Psi ^{Br} _{b} \rangle$ and $ | \Psi ^{Br} _{j} \rangle$ by the ionic-covalent form, and approximate the one- and three-electron states of the one-bond system by $| \Psi ^{\prime} _{b,1,\sigma} \rangle $ and $ | \Psi ^{\prime} _{b,3,\sigma} \rangle $, which are introduced in Appendix B. Let $| \Psi ^{ \prime (j)} _{b,1,\sigma} \rangle = c _{j A^{\prime} \sigma } ^{\dagger} |0 \rangle$ and $| \Psi ^{\prime (j)} _{b,3,\sigma} \rangle = c _{j B^{\prime} \sigma } | Fb ^{(j)} \rangle $ as the $j$-th bond's states mapped from $| \Psi ^{ \prime } _{b,1,\sigma} \rangle$ and $| \Psi ^{ \prime} _{b,3,\sigma} \rangle$. Here the 4-electron state $|Fb ^{(j)} \rangle$ describes the fully occupied $j$-th bond, and the annihilators $c _{j A^{\prime} \sigma } = \sqrt{ 1 - | \lambda _{3} | ^{2} } c _{ j A \sigma } + \lambda _{3} c _{ j B \sigma }$ and $c _{j B^{\prime} \sigma } = - \lambda _{3} ^{*} c _{ j A \sigma } + \sqrt{ 1 - | \lambda _{3} | ^{2} } c _{ j B \sigma }$. The effective SCF state for the compound is $| \Psi ^{SCF} _{Cr}  \rangle = \prod _{j=1} ^{ N }  c _{ j A^{\prime \prime} \uparrow } ^ {\dag} c _{j A^{\prime \prime} \downarrow } ^ {\dag} | 0 \rangle$, and $E _{Cr} ^{SCF} =\langle \Psi ^{SCF} _{Cr} | {\cal H} | \Psi ^{SCF} _{Cr}  \rangle$ is the SCF value of $E _{Cr}$. 

For the electron affinity and ionization discussed in subsection IV-B, in the SCF calculation \cite{Stoyanova,Grafenstein2} we shall calculate $ \langle \Psi ^{SCF}  _{ \eta ^{\prime} \sigma, + } | {\cal H} | \Psi ^{SCF}  _{ \eta ^{\prime} \sigma, + }  \rangle$ and $\langle \Psi ^{SCF}  _{ \eta ^{\prime} \sigma, - } | {\cal H} | \Psi ^{SCF}  _{ \eta ^{\prime} \sigma, - }  \rangle$ when the carrier excitations occur in the subsystem corresponding to $| \eta ^{\prime} \rangle$. Here
\begin{eqnarray}
| \Psi ^{SCF} _{ \eta ^{\prime} \sigma, \pm }  \rangle= 
\begin{cases}
\sum _{j=1} ^{N} \langle w _{j} | \eta ^{ \prime} \rangle c _{ j B^{ \prime \prime} \underline{ \sigma } } ^{\dag} | \Psi ^{SCF} _{Cr} \rangle \text{ \ \ for \ \ } + \\
\sum _{j=1} ^{N}  \langle w _{j} | \eta ^{ \prime} \rangle ^{*} c _{ j A^{ \prime \prime} \underline{ \sigma } } | \Psi ^{SCF} _{Cr}  \rangle \text{ \  for \ \ } -
\end{cases}
\end{eqnarray}
with $\underline{\sigma} = - \sigma$. We can denote the spin-independent values $ \langle \Psi ^{SCF}  _{ \eta ^{\prime} \sigma, + } | {\cal H} | \Psi ^{SCF}  _{ \eta ^{\prime} \sigma, + }  \rangle$ and $\langle \Psi ^{SCF}  _{ \eta ^{\prime} \sigma , - } | {\cal H} | \Psi ^{SCF}  _{ \eta ^{\prime} \sigma , - }  \rangle$ as $E ^{SCF} _{ \eta ^{\prime} , +} $ and $E ^{SCF} _{ \eta ^{\prime} , -} $, respectively, and choose $\sigma= \downarrow$ in Eq. (26) without loss of generality. In the SCF calculation, the added/removed charge in the $j$-th bond equals $| \langle w _{j} | \eta ^{\prime} \rangle |^{2}$ just as that in subsection IV-B, and $E ^{ ( \eta ^{\prime} ,\pm ) }_{Cr} - E _{Cr}$ is approximated as 
\begin{eqnarray}
E ^{SCF} _{ \eta ^{\prime} , \pm} - E ^{SCF}_{Cr} = {\cal K} ^{SCF} _{ \eta ^{\prime} , \pm} + {\cal B}^{SCF} _{ \eta ^{\prime} , \pm} \text{ \ with } 
\end{eqnarray}  
\[  
\begin{cases}
{\cal B}^{SCF} _{ \eta ^{\prime} , \pm} = \sum _{j=1} ^{ N } | \langle w _{j} | \eta ^{\prime} \rangle |^{2} [ \langle j , \pm| (H_{sb} ^{(j)} + \hat{ U } _{j})|  j , \pm \rangle - \langle  \Psi ^{SCF}_{j} |  (H_{sb} ^{(j)} + \hat{ U } _{j}) |  \Psi ^{SCF} _{j} \rangle ] \\
{\cal K} ^{SCF} _{ \eta ^{\prime} , +} = \sum _{ j \neq j ^{\prime} } \sum _{ \xi \xi ^{\prime} }  t _{ j \xi , j ^{\prime} \xi ^{\prime} } \langle w _{ j ^{\prime} } | \eta ^{\prime} \rangle \langle \eta ^{ \prime} | w _{ j } \rangle \langle j ,+| c _{j \xi \uparrow } ^{\dag} | \Psi ^{SCF} _{j} \rangle \langle  \Psi ^{SCF} _{j ^{\prime} } |  c _{j ^{\prime} \xi ^{\prime} \uparrow } | j ^{\prime} , + \rangle \\
{\cal K} ^{SCF} _{ \eta ^{\prime} , -} = - \sum _{ j \neq j ^{\prime} } \sum _{ \xi \xi ^{\prime} }  t _{ j \xi , j ^{\prime} \xi ^{\prime} } \langle w _{ j ^{\prime} } | \eta ^{\prime} \rangle \langle \eta ^{ \prime} | w _{ j } \rangle \langle \Psi ^{SCF} _{j}| c _{j \xi \uparrow } ^{\dag} | j , - \rangle \langle  j ^{\prime} , - |  c _{j ^{\prime} \xi ^{\prime} \uparrow } |\Psi ^{SCF} _{j ^{\prime} }  \rangle
\end{cases}.
\]
Here $| j , - \rangle = c _{j A^{\prime \prime} \uparrow } | \Psi ^{SCF} _{j} \rangle$ and $| j , + \rangle = c _{j B^{\prime \prime} \uparrow } ^{\dag} | \Psi ^{SCF} _{j} \rangle$. The energy factors ${\cal B}^{SCF} _{ \eta ^{\prime} , \pm}$ and ${\cal K} ^{SCF} _{ \eta ^{\prime} , \pm}$ are due to $\sum _{ j } (H_{sb} ^{(j)} + \hat{U _{j} })$ and $ \sum _{ j \neq j ^{ \prime}} H_{hop} ^{(j, j^{\prime})} $, respectively.  

To include the coupled-cluster corrections, in Eq. (27) we shall consider the two-particle excitation \cite{Stoyanova,Grafenstein2} ${\cal S}^{(2)}_{J}= \frac{\lambda _{2}}{ \sqrt{ 1 - |\lambda _{2}| ^{2} } } c _{J B^{\prime \prime} \uparrow } ^ {\dag}  c _{ J A^{\prime \prime} \uparrow } c _{J B^{\prime \prime} \downarrow } ^ {\dag}  c _{J A^{\prime \prime} \downarrow } $ to replace $| \Psi ^{SCF} _{J} \rangle $ by $ | \Psi  _{J} ^{Br} \rangle = \sqrt{ 1 - |\lambda _{2}| ^{2} } \times exp( {\cal S} ^{(2)} _{J}) | \Psi ^{SCF} _{J} \rangle $ for any $J=j$ or $j ^{\prime}$. In addition, we may introduce the one-particle excitation \cite{Stoyanova,Grafenstein2} ${\cal S}^{(1)}_{J}= \frac{\lambda _{4}}{ \sqrt{ 1 - |\lambda _{4}| ^{2} } } c _{ J B ^{\prime \prime} \downarrow } ^{ \dag } c _{J A^{\prime \prime} \downarrow }  $ to modify $| J, - \rangle$ and $| J, + \rangle$ as $| \Psi ^{\prime} _{b,1,\downarrow} \rangle = \sqrt{ 1 - |\lambda _{4}| ^{2} }  \times exp ( {\cal S}^{(1)}_{J} ) | J , - \rangle $ and $| \Psi ^{\prime} _{b,3,\downarrow} \rangle= \sqrt{ 1 - |\lambda _{4}| ^{2} }  \times exp( {\cal S}^{(1)}_{J} ) | J , + \rangle $, respectively, in the $J$-th bond. Here $\lambda _{4} = \lambda _{3} ^{\ast} \sqrt{ 1 - | \lambda _{1} | ^{2} } - \lambda _{1} \sqrt{ 1 - | \lambda _{3} | ^{2} } $. The generator for the ground-state correlation \cite{Grafenstein} equals $\sum _{J} {\cal S} ^{(2)} _{J}$, and the operator ${\cal S} ^{(1)} _{j} + \sum _{j \neq J} {\cal S} ^{(2)} _{J}$ generates the correlated states corresponding to $c _{ j B^{ \prime \prime} \uparrow } ^{\dag} | \Psi ^{SCF} _{Cr} \rangle$ and $c _{ j A^{ \prime \prime} \uparrow } | \Psi ^{SCF} _{Cr} \rangle$ \cite{Grafenstein,Stoyanova,Grafenstein2}.

When the coupled-cluster method is applicable near the ionic limit, as mentioned in Appendix B, it is important to perform the orbital transformation in Eq. (25) to improve my model. Rewriting $| \Psi  _{b} ^{Br} \rangle $ as the ionic-covalent form by performing such a transformation, in the improved model we can re-obtain the difference $ E _{b} ^{(\pm)} - E _{b}$ in Eq. (22) by using the coupled-cluster method to modify ${\cal B}^{SCF} _{ \eta ^{\prime} , \pm}$. The coupled-cluster corrections to $ {\cal K} ^{SCF} _{ \eta ^{\prime} , \pm}$, in fact, are small and we have $\langle \eta ^{ \prime} | H _{\pm } | \eta ^{ \prime } \rangle \simeq {\cal K} ^{SCF} _{ \eta ^{\prime} , \pm}$ under the assumption about the small hopping coefficients. So we can use the coupled-cluster method to correct $E ^{SCF} _{ \eta ^{\prime} , \pm} - E ^{SCF}_{Cr}$ and obtain the difference close to $E_{Cr} ^{ ( \eta ^{\prime} , \pm ) } -E_{Cr}$. In addition, the added/removed charge in the $j$-th bond equals $| \langle w _{j} | \eta ^{\prime} \rangle |^{2}$, which is the same as that in subsection IV-B, because the above one- and two-particle excitations do not change the number of electrons in each bond. Therefore, my improved model can be supported by the coupled-cluster theory near the ionic limit.  

\section*{Appendix D}

The matrix $\rho ^{(2)} _{ \omega _{j} }$ in Eq. (14) represents the quasielectron at the orbital $ | \omega _{j} \rangle \otimes | L  \rangle = \alpha _{i} | A, j \rangle + \alpha _{c} | B, j \rangle$. When the ionic-covalent bonds in the compound are not identical to each other, the bonding coefficients can depend on $j$ and we shall replace $\alpha _{i}$ and $\alpha _{c}$ by $\alpha _{i} ^{(j)}$ and $\alpha _{c} ^{(j)}$ in the $j$-th bond. Here the coefficients $\alpha _{i} ^{(j)}$ and $\alpha _{c} ^{(j)}$ satisfy $| \alpha _{i} ^{(j)} | ^{2} + | \alpha _{c} ^{(j)} | ^{2} =1$ for all $j=1 \sim N$. The matrix $\rho ^{(2)} _{ \omega _{j} }$ should be modified as $(\alpha _{i} ^{(j)} | A, j \rangle + \alpha _{c} ^{(j)} | B, j \rangle ) (\alpha _{i} ^{(j)} \langle A, j | + \alpha _{c} ^{(j)} \langle B, j |) $ and we can still take $\rho ^{(1)} _{ \omega _{j} } = | A, j \rangle \langle A, j |$. Equations (14) and (15) remain valid after the modification, and the energy $E _{Cr}$ can still be obtained based on Eq. (18). 

In subsection IV-B, the quasielectron at $| \eta ^{\prime } \rangle \otimes | L \rangle = \sum _{j} \langle \omega _{j} | \eta ^{\prime} \rangle (\alpha _{i} | A, j \rangle + \alpha _{c} | B, j \rangle)$ is ionized from subsystem $\eta ^{ \prime}$ while one quasielectron enters $| \eta ^{\prime } \rangle \otimes | \bar{L} \rangle  = \sum _{j} \langle \omega _{j} | \eta ^{\prime} \rangle (\alpha _{c} ^{*} | A, j \rangle - \alpha _{i} ^{*} | B, j \rangle)$ in the affinitive process. When the coefficients for the ionic and covalent parts depend on $j$, the orbital  of the quasielectron to be ionized should be modified as $ \sum _{j} \langle \omega _{j} | \eta ^{\prime} \rangle (\alpha _{i} ^{(j)} | A, j \rangle + \alpha _{c} ^{(j)} | B, j \rangle) \equiv| \eta ^{\prime} _{L} \rangle $. In addition, the orbital for the added quasielectron becomes $ \sum _{j} \langle \omega _{j} | \eta ^{\prime} \rangle (\alpha _{c} ^{(j) \ast} | A, j \rangle - \alpha _{i} ^{(j) \ast} | B, j \rangle) \equiv | \eta ^{\prime} _{ \bar{L} } \rangle$. So we need to modify the first two matrices in Eq. (21) as $\rho ^{( \text{II} )} _{ \eta ^{\prime} , + }  = \rho ^{( \text{II} )} + | \eta ^{\prime} _{ \bar{L} } \rangle \langle \eta ^{\prime} _{ \bar{L} } |$ and $\rho ^{( \text{II} )} _{ \eta ^{\prime} , - } = \rho ^{( \text{II} )} - | \eta ^{\prime} _{L} \rangle \langle \eta ^{\prime} _{L} |$. The energy $E_{Cr} ^{ ( \eta ^{\prime} , \pm ) } $ can be calculated based on Eq. (20), where $d ^{( \text{I} )} _{ \eta ^{\prime} ,\pm } $ and $d ^{( \text{II} )} _{ \eta ^{\prime} ,\pm } $ can still be obtained from the last two lines in Eq. (21). In the compound composed of different chemical bonds, therefore, Eqs. (18) and (20) may yield the difference $E_{Cr} ^{ ( \eta ^{\prime} , \pm ) } - E_{Cr}$ for the electron ionization and affinity under the suitable modification.

\begin{figure}
\begin{center}
\includegraphics[width=5in]{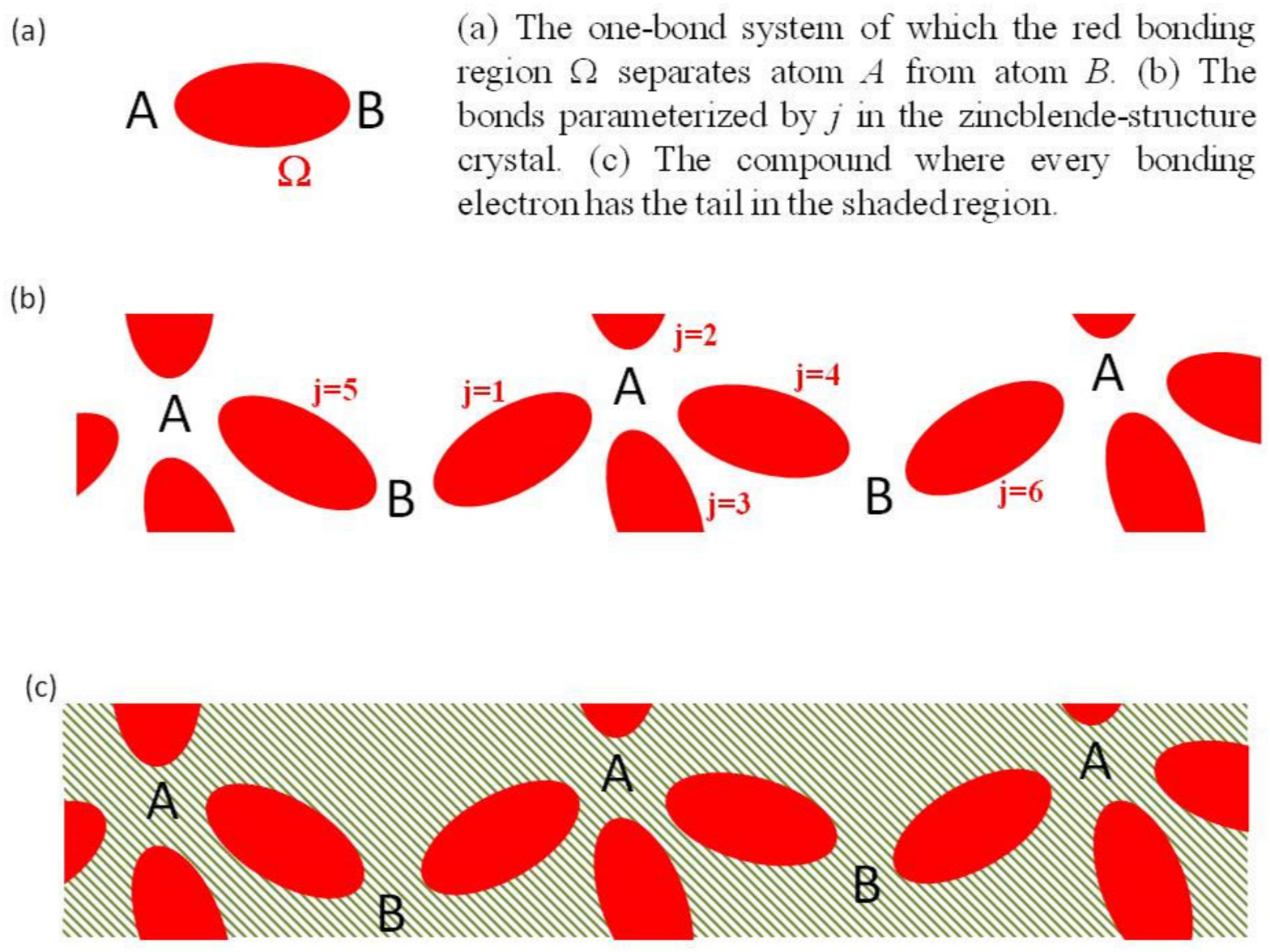}
\caption{The bonding systems.}
\end{center}
\end{figure}
\begin{figure}
\begin{center}
\includegraphics[width=5in]{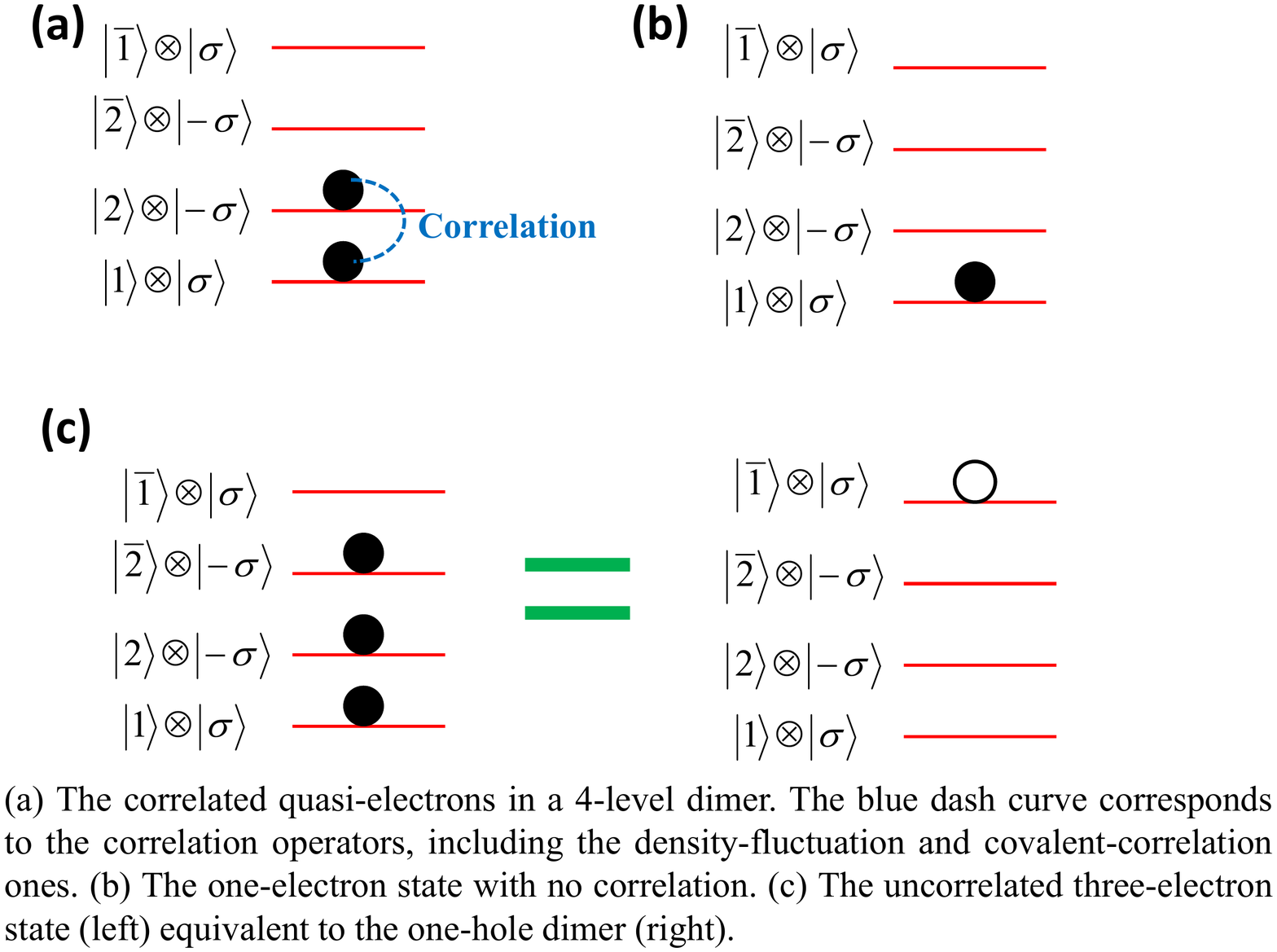}
\caption{The 4-level dimer.}
\end{center}
\end{figure}
\begin{figure}
\begin{center}
\includegraphics[width=5in]{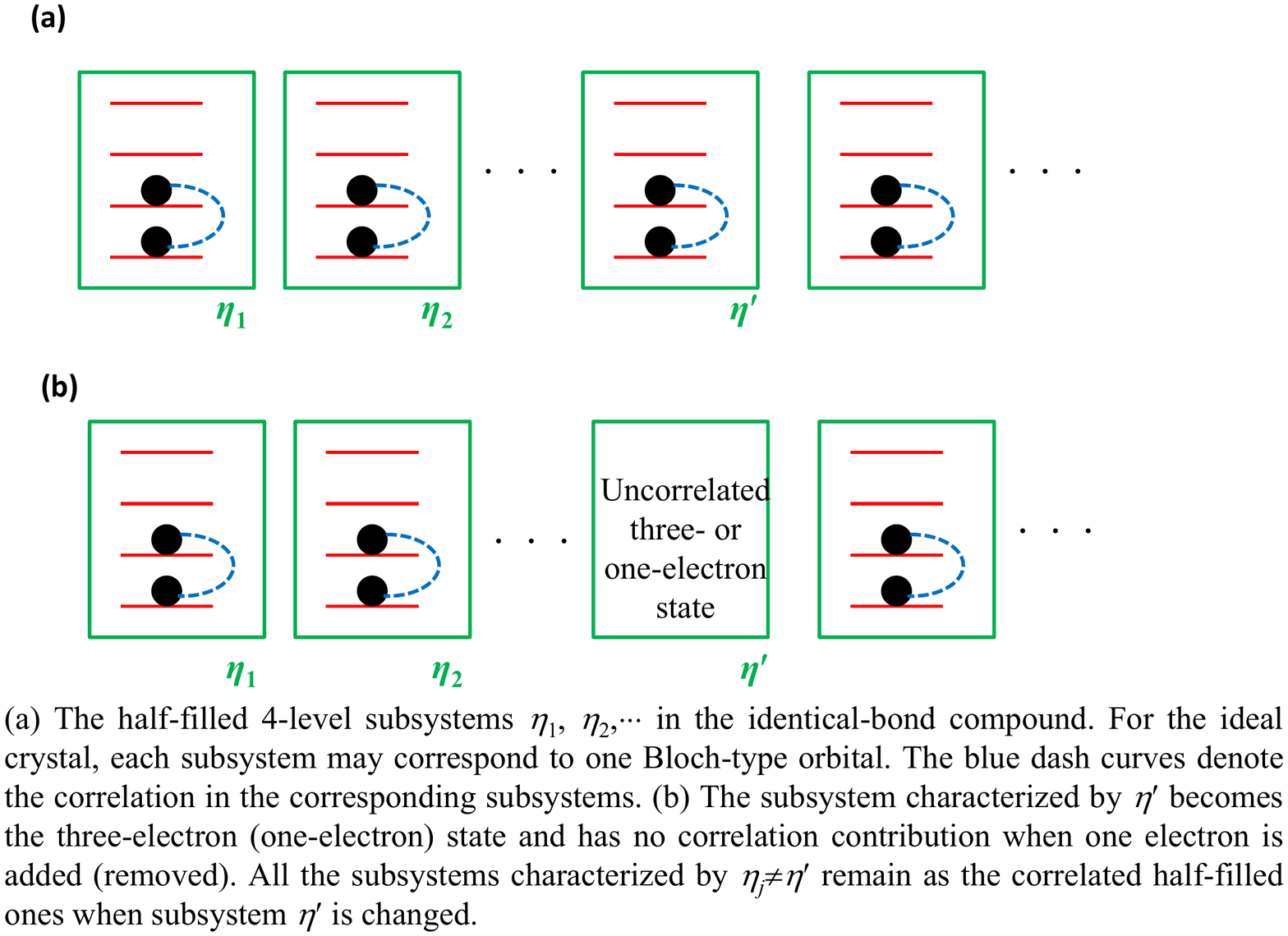}
\caption{The 4-level subsystems in the compound system composed of identical ionic-covalent dimers.}
\end{center}
\end{figure}
\begin{figure}
\begin{center}
\includegraphics[width=5in]{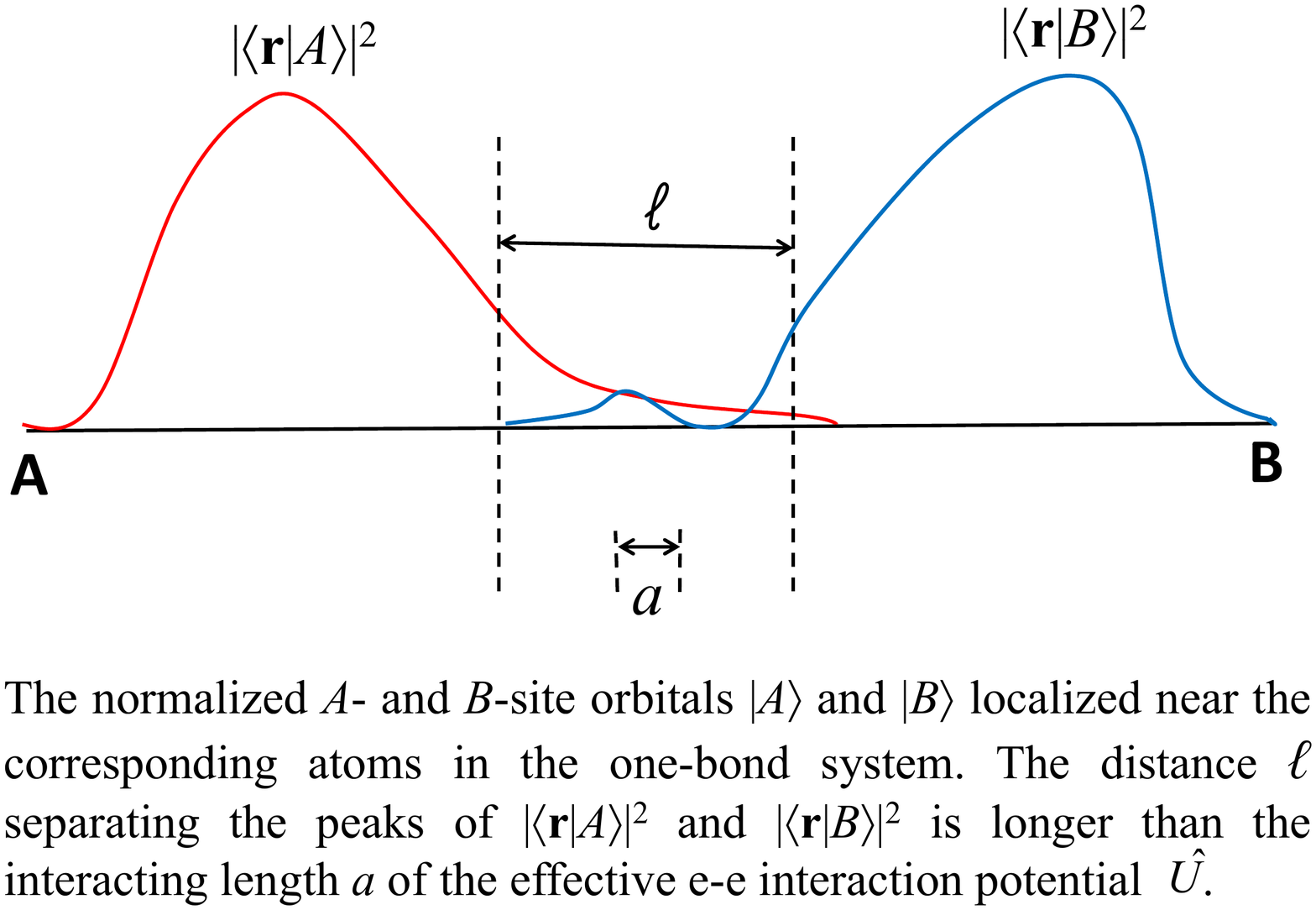}
\caption{The site orbitals in the one-bond system.}
\end{center}
\end{figure}
\end{document}